\documentclass[iop]{emulateapj}
\usepackage[latin1]{inputenc}
\usepackage[english]{babel}
\usepackage{amsmath}
\usepackage{amssymb}
\usepackage{latexsym}
\usepackage{epsfig}
\usepackage{subfigure}
\usepackage{natbib}
\usepackage{setspace}
\usepackage{graphicx}
\usepackage{longtable}
\usepackage{pdfpages}
\DeclareGraphicsExtensions{.pdf}

\shorttitle{Stellar rotation effects in polarimetric microlensing}
\shortauthors{Sajadian}
\begin{document}

\title{Stellar rotation effects in polarimetric microlensing}
\author{Sedighe Sajadian \altaffilmark{1}}
\altaffiltext{1}{Department of Physics, Sharif University of
Technology, P.O. Box 11155-9161, Tehran, Iran
}\email{sajadian@ipm.ir}

\begin{abstract}
It is well known that the polarization signal in microlensing events
of hot stars is larger than that of main-sequence stars. Most hot
stars rapidly rotate around their stellar axes. The stellar rotation
makes ellipticity and gravity-darkening effects which break the
spherical symmetry of the source shape and the circular symmetry of
the source surface brightness respectively. Hence, it causes a net
polarization signal for the source star. This polarization signal
should be considered in polarimetry microlensing of fast rotating
stars. For moderate rotating stars, lensing can magnify or even
characterize small polarization signals due to the stellar rotation
through polarimetry observations. The gravity-darkening effect due
to a rotating source star makes asymmetric perturbations in
polarimetry and photometry microlensing curves whose maximum happens
when the lens trajectory crosses the projected position of the
rotation pole on the sky plane. The stellar ellipticity makes a time
shift (i) in the position of the second peak of the polarimetry
curves in transit microlensing events and (ii) in the peak position
of the polarimetry curves with respect to the photometry peak
position in bypass microlensing events. By measuring this time shift
via polarimetry observations of microlensing events, we can evaluate
the ellipticity of the projected source surface on the sky plane.
Given the characterizations of the FOcal Reducer and low dispersion
Spectrograph (FORS2) polarimeter at Very Large Telescope (VLT)
telescope, the probability of observing this time shift is so small.
The more accurate polarimeters of the next generation may likely
measure these time shifts and evaluate the ellipticity of microlensing source stars.\\
\end{abstract}

\section{Introduction}
Stellar rotation refers to the angular motion of a star around its
axis. If there is no stellar rotation, the gravitational force
condenses celestial bodes into perfect spheres. Whereas if an object
rotates around its axis, some portion of the gravitational
attraction provides the centrifugal acceleration whose value depends
on the stellar latitude and decreases by increasing it from the
stellar equator to the stellar pole. Therefore, the stellar rotation
creates stellar oblateness \cite{Collins66,Lebovitz1967}. For
rapidly rotating stars the polar surface brightness is more than the
equatorial one, the so-called gravity-darkening effect. This effect
is resulted from the von Zeipel (1924)'s theorem, i.e. the radiative
flux is proportional to the local effective gravity.

For main-sequence stars the stellar angular velocity
$\mathrm{\Omega}$ decreases with the stellar age $\mathrm{t}$:
$\mathrm{\Omega~\propto~t^{-1/2}}$, i.e. the Skumanich's
relationship \cite{skumanich1972,Durney1978}. According to this
relation, the star age can be derived by the rotational rate
\cite{Barnes2007}. Another result of this relation is that the
rotational velocities of pre-main sequence stars are higher than
those of main sequence stars. The stars in the spectral class
between $\mathrm{F5}$ up to $\mathrm{O5}$ often rotate fast with the
mean rotational velocity of order of $\mathrm{100-200~km~s^{-1}}$
\cite{McAlister2005,Peterson2004}. Also, their rotational velocity
increases by mass and it maximizes for massive $\mathrm{B}$-class
stars. The less massive stars have much lower rotational speeds
about a few $\mathrm{km~s^{-1}}$ after a few $\mathrm{100~Myr}$
\cite{Kraft1970,Gallet2013}, since the magnetized stellar winds over
the surface of these stars transport the angular momentum
\cite{Schatzman62}. However, Irwin et al. (2011) by measuring the
rotational period of stars with masses less than $\mathrm{0.35
M_{\odot}}$ found some exceptionally fast and slow rotators. Also,
brown dwarfs have the averaged period of order of $\mathrm{15}$
hours at young ages (Scholz \& Eisl\"{o}ffel 2004, 2005).

The stellar rotation is determined via several methods. The
projected radial velocity of rotating stars can be measured by
spectroscopy observations of the Doppler broadening in the
absorption lines of stars \cite{Abney1877}. However, this method is
not convenient for slowly rotating stars with the projected radial
velocity less than $\mathrm{20~km~s^{-1}}$ \cite{Bouvier2013}. For
nearby fast rotating stars, the stellar oblateness projected on the
sky plane can be measured by the interferometry method, e.g.
Kervella et al. (2004). Combining the spectroscopy and
interferometry methods helps us to charcterize the inclination angle
of rotational axis with respect to the sky plane, e.g. Le Bouquin et
al. (2009). The rotational periods of nearby spotted stars can also
be determined via photometry observations of their light curves,
e.g. Affer et al. (2012) and McQuillan et al. (2013). In that case,
the stellar spots on rotating stars disturb the star light curves
periodically. This method was first used to measure the Sun
rotational period by Galileo Galilei \cite{Casas2006}. Generally,
the mentioned methods can identify the rotational properties of
nearby or bright (and massive) stars.

Gravitational microlensing is a useful tool to highlight the
rotational properties of distant stars. One channel is photometry
observations of these events. In that case, the ellipticity or the
radial velocity of source stars can perturb microlensing light
curves, some aspects of this subject were studied by a number of
authors. Heyrovsk\'y \& Loeb (1997) introduced an efficient method
for calculating the microlensing light curve of an elliptical source
by a point-mass lens and studied some properties of these light
curves. The projected radial velocity of source stars or even the
Einstein angular radius can be measured by spectroscopic
observations of high-magnification microlensing events
\cite{Maoz1994,Gould1997}. Also Gaudi \& Haiman (2004) studied
microlensing of elliptical sources by fold caustics and concluded
that the ellipticity deviation in microlensing light curves is
qualitatively similar to that due to the limb-darkening effect. In
these references, the gravity-darkening effect resulted from the
stellar ellipticity has not been considered.

In this work, we study the possibility to detect and characterize
the rotational properties of distant source stars by performing
polarimetry observations of high-magnification microlensing events.
Rotating stars are oblate objects and their projected surface
brightness will not be symmetric owing to the projection process
unless their rotation axis orients toward the observer.
Consequently, a rotating star has a net polarization signal whose
value is a function of the stellar angular velocity and the
inclination angle of its rotation axis. Lensing can magnify these
small polarization signals and cause them to be realized or even
characterized through polarimetry observations.

In section (\ref{three}), we explain the formalism used for
calculating the polarization signal of an elliptical source star.
The properties of polarimetry microlensing events of elliptical
sources will be studied in the next section. In section
(\ref{five}), we first study the statistic of fast rotating stars
using the Kepler data. Then, we do a Monte Carlo simulation of
high-magnification microlensing events of rotating stars toward the
Galactic spiral arms to evaluate the efficiency for detecting the
rotation-induced (polarimetry and photometry) perturbations of
source stars. Finally, we conclude in the last section.

\section{The polarization signal of an elliptical source}\label{three}
The intrinsic polarization signal of rapidly rotating early-type
stars was first studied by Harrington \& Collins (1968). Also, the
scattering polarization generated by nonradial pulsating stars was
calculated by Stamford \& Watson (1980). Bjorkman \& Bjorkman (1994)
have analytically calculated the polarization signal due to a fast
rotating star surrounded by an axisymmetric disk. Al-Malki et al.
(1999) and Ignace et al. (2009) estimated the Stokes parameters due
to anisotropic light sources with spherical envelopes and envelopes
of arbitrary shapes by ignoring the finite size effect of the source
star. Recently, several codes were written to solve the coupled
problem of the Non-Local Thermodynamic Equilibrium (NLTE) and
radiative equilibrium for arbitrary three-dimensional envelope
geometries, using the Monte Carlo method ( e.g. Carciofi \& Bjorkman
2006, Carciofi et al. 2006, Whitney et al. 2013).

We use the formalism developed by Harrington \& Collins (1968) to
calculate the intrinsic polarization signals of rotating stars.
Accordingly, we assume that (i) The stellar envelope is optically
thin enough to use the single scattering approximation. (ii) The
rotational angular velocity is constant over the stellar surface and
there is no differential rotation effect (see e.g. Kitchatinov
2005). (iii) The stellar angular momentum is not transported from
the stellar surface during microlensing events. (iv) The star is
rotating as a rigid body so that the von Zeipel's theorem is
applicable. (v) The magnetic field of the source star is negligible.
(vi) There is no disk around the source star. (vii) The angular
rotation velocity normalized to the break-up velocity of the source
star is small enough to approximate the local polarization over the
rotating source surface with the local polarization over a spherical
source surface. Note that fast rotating stars, e.g.
$\mathrm{B}$-class stars, are intrinsically variable and are not
suitable candidates to be studied for microlensing observations.

We consider some parameters to describe an elliptical source: (i)
the stellar equatorial and polar radii $\mathrm{R_{eq}}$ and
$\mathrm{R_{p}}$. (ii) The inclination angle of the stellar rotation
axis with respect to the sky plane $\mathrm{i}$. (iii) Two
parameters to represent the limb-darkening coefficients
$\mathrm{c_{1}}$ and $\mathrm{c_{2}}$.

To describe an elliptical source in the sky plane we need two
coordinate systems: (i) Observer coordinate frame
$\mathrm{(x_{o},y_{o},z_{o})}$ so that the projected source center
is at its origin, the observer is on the $\mathrm{z_{o}}$-axis at
$\mathrm{+\infty}$ and the $\mathrm{z_{o}-y_{o}}$ plane contains the
stellar rotation axis (i.e. its $\mathrm{x_{o}}$-axis is parallel
with the semimajor axis of the projected elliptical source). (ii)
Stellar coordinate system $\mathrm{(x_{\star},y_{\star},z_{\star})}$
so that the $\mathrm{y_{\star}}$-axis is along the stellar rotation
axis and the $\mathrm{x_{\star}}$-axis is along the observer's
$\mathrm{x_{o}}$-axis. We transform the second coordinate system to
the first one by a rotation around $\mathrm{x_{\star}}$-axis by the
inclination angle $\mathrm{-i^{\circ}}$, so that:
\begin{eqnarray}\label{eq1}
x_{o}&=&x_{\star},\nonumber\\
y_{o}&=&y_{\star} \cos(i)- z_{\star}\sin(i),\nonumber\\
z_{o}&=&y_{\star}\sin(i)+ z_{\star}\cos(i).
\end{eqnarray}
We use $\mathrm{(R_{\star},\theta_{\star},\phi_{\star})}$ to
represent points over the stellar surface in the spherical stellar
coordinate, i.e.
\begin{eqnarray}
x_{\star}&=&R_{eq}\sin(\theta_{\star})\sin(\phi_{\star}),\nonumber\\
y_{\star}&=&R_{p}\cos(\theta_{\star}),\nonumber\\
z_{\star}&=&R_{eq}\sin(\theta_{\star})\cos(\phi_{\star}).
\end{eqnarray}
Noting that $\mathrm{R_{eq}}$ can be determined according to the
stellar angular velocity $\mathrm{\Omega}$. Generally,
$\mathrm{R_{\star}=\sqrt{x_{\star}^{2}+y_{\star}^{2}+z_{\star}^{2}}}$
depends on $\mathrm{\theta_{\star}}$ owing to the stellar oblateness
\cite{Collins66}:
\begin{eqnarray}\label{Rstar}
R_{\star}(\omega,\theta_{\star})=\frac{3R_{p}}{w
\sin\theta_{\star}}\cos[\frac{\pi+\cos^{-1}
(\omega~\sin\theta_{\star})}{3}],
\end{eqnarray}
where $\mathrm{\omega=\Omega/\Omega_{crit}}$ is the ratio of the
star's angular velocity to the critical or break-up velocity
$\mathrm{\Omega_{crit}}$ at which the centrifugal force at the
stellar equator becomes equal to the gravitational attraction, given
by: $\mathrm{\Omega^{2}_{cirt}=(2/3)^{3}G~M_{\star}/R^{3}_{p}}$
where $\mathrm{M_{\star}}$ is the stellar mass and
$\mathrm{\Omega_{crit}}$ is measured in the unit of radian per
second.

In our formalism, the projected position of the stellar rotation
pole in the observer coordinate system is
$\mathrm{\boldsymbol{d}_{R}^{\dag}=(0,R_{p}\cos(i),0)}$. The
emergent radiative flux of the source star in this point maximizes.
The projection of the source star on the sky plane is an ellipse
whose semimajor and minor axes are $\mathrm{a=\mathrm{R_{eq}}}$ and
$\mathrm{b=\sqrt{R_{eq}^{2}\sin^{2}(i) + R_{p}^{2}\cos^{2}(i)}}$
respectively. Also the $\mathrm{x_{o}}$ and $\mathrm{y_{o}}$-axes
are toward its semimajor and minor axes. We use
$\mathrm{(\rho,\phi)}$ to represent points over the stellar surface
projected on the sky plane in the polar observer coordinate where
$\mathrm{\phi\in[-\pi,\pi]}$ and $\mathrm{\rho\in[0,\rho_{m}]}$.
$\mathrm{\rho_{m}=b/\sqrt{(b\cos\phi)^{2}+(a\sin\phi)^{2}}}$ is
normalized to $\mathrm{a}$ and in the range of
$\mathrm{\rho_{m}\in[b/a,1]}$.


To calculate the polarization signal of an elliptical star, we use
the Stokes intensities. There are four Stokes intensities
$\mathrm{I_{I}}$, $\mathrm{I_{Q}}$, $\mathrm{I_{U}}$ and
$\mathrm{I_{V}}$ which represent the total intensity, two components
of linear polarized intensities and circular polarized intensity
over the source surface, respectively \cite{Tinbergen96}. Taking
into account that there is only linear polarization of light
scattered on the stellar atmosphere, we set $\mathrm{I_{V}=0}$. The
other intensities are estimated according to the limb-darkening
effect producing a local polarization owing to the light scattering
in the stellar atmosphere. This local polarization depends on the
scatterer species and the nature of the source star (Ingrosso et al.
2012,2015). For example, in late-type main-sequence stars, the
polarization signal is generated rather owing to Rayleigh scattering
on neutral hydrogens and a bit due to Thompson scattering by free
electrons \cite{Fluri1999}. In cool giant stars Rayleigh scattering
on atomic and molecular species or on dust grains generates the
polarization signal. In that case, Simmons et al. (2002) introduced
the appropriate Stokes parameters for giant stars with spherical
circumstellar envelopes lensed by a single lens. In hot early-type
stars with a free electron atmosphere mostly Thompson scattering
produces the polarization signal \cite{Chandrasekhar60}. The
magnitude of Stokes (total and polarized) intensities over the
surface of these stars can be written in the form
\cite{schneider87}:
\begin{eqnarray}\label{II}
I_{I}(\rho,\phi)&=&I_{0,o}(\rho,\phi)~[1-c_{1}(1-\mu)],\nonumber\\
I_{P}(\rho,\phi)&=&I_{0,o}(\rho,\phi)~[c_{2}(1-\mu)],
\end{eqnarray}
where $\mathrm{I_{P}=\sqrt{I_{Q}^{2}+I_{U}^{2}}}$,
$\mathrm{c_{1}=0.64}$, $\mathrm{c_{2}=0.032}$,
$\mathrm{\mu=\sqrt{1-\rho^{2}/\rho_{m}^{2}}}$ and
$\mathrm{I_{0,o}(\rho,\phi)}$ is the emergent radiative flux of the
source star in the observer coordinate system. Note that
Chandrasekhar (1960) considered a spherically symmetric,
isotropically scattering atmosphere to calculate the Stokes
intensities. Indeed, we assume that the effect of the stellar
rotation is small enough to adopt for rotating stars the local
Stokes intensities in equation (\ref{II}). We calculate the
radiative flux $\mathrm{I_{0,o}}$ in the observer coordinate system
starting from the flux in the stellar coordinate system
$\mathrm{I_{0,s}}$ and using the coordinate transformations in
equation (\ref{eq1}). The gravity-darkening effect which is a result
of the stellar ellipticity causes the intrinsic radiative flux of
the source star $\mathrm{I_{0,s}}$ changes over the source surface.
According to the von Zeipel (1924)'s theorem, we can express the
stellar radiative flux as a function of the effective gravity
$\mathrm{g_{eff}}$ over the surface of a uniformly rotating star (in
the stellar coordinate) which is given by:
\begin{eqnarray}
I_{0,s}(\Omega,\theta_{\star})=-\frac{L_{\star}(\mathcal{P})}{4~\pi~G~M_{\star}(\mathcal{P})}|\boldsymbol{g}_{eff}(\Omega,\theta_{\star})|,
\end{eqnarray}
where the stellar luminosity $\mathrm{L_{\star}}$ and the stellar
mass $\mathrm{M_{\star}}$ are evaluated on a surface of a constant
pressure $\mathrm{\mathcal{P}}$. If there is no angular momentum
transport on the star surface, the vector of the effective gravity
in the Roche model and in the hydrostatic equilibrium is given by
(see e.g. Maeder \& Meynet 2011):
\begin{eqnarray}\label{geff}
\boldsymbol{g}_{eff}(\Omega,\theta_{\star})&=&[-\frac{GM_{\star}}{R_{\star}^{2}(\theta_{\star})}+\Omega^{2}R_{\star}(\theta_{\star})\sin^{2}\theta_{\star}]\boldsymbol{e}_{r}\nonumber\\
&+&[\Omega^{2}R_{\star}(\theta_{\star})\sin\theta_{\star}\cos\theta_{\star}]\boldsymbol{e}_{\theta},
\end{eqnarray}
where $\mathrm{\boldsymbol{e}_{r}}$ and
$\mathrm{\boldsymbol{e}_{\theta}}$ are the unit vectors in the
radial and latitudinal directions. Accordingly, $\mathrm{I_{0,s}}$
for an elliptical source is proportional to
$\mathrm{|\boldsymbol{g}_{eff}(\Omega,\theta_{\star})|}$ and a
function of $\mathrm{(\Omega,\theta_{\star})}$.


Also we assume that the stellar atmosphere has an elliptical shape
due to the stellar rotation, so that polarization vectors are
tangential to co-center ellipses whose centers coincide to the
stellar center. The normal vector to the elliptical source surface
in each point $\mathrm{(x_{o},y_{o})}$ from the source center is
given by: $\mathrm{\boldsymbol{n}=(x_{o}/a^{2},y_{o}/b^{2})}$.

Now by integrating these Stokes intensities over the source surface,
the corresponding Stokes parameters $\mathrm{S_{I}}$,
$\mathrm{S_{Q}}$ and $\mathrm{S_{U}}$ are obtained. If the source
light is magnified by a microlens, we should add a weight function
for each source surface element, i.e. the magnification factor
$\mathrm{A}$. In that case, the Stokes parameters are given by:
\begin{eqnarray}\label{tsparam}
S_{I}&=&~\rho^2_{\star}\int_{-\pi}^{\pi}d\phi\int_{0}^{\rho_{m}}\rho~d\rho I_{I}(\rho,\phi)~ A(u),\\
\left( \begin{array}{c} S_{Q}\\
S_{U}\end{array}\right)&=&\rho^2_{\star}\int_{-\pi}^{\pi}d\phi\int_{0}^{\rho_{m}}\rho~d\rho I_{P}(\rho,\phi) A(u) \left( \begin{array}{c} -\cos 2\varphi \nonumber \\
-\sin 2\varphi \end{array} \right),
\end{eqnarray}
where $\mathrm{\rho_{\star}=R_{eq}x_{rel}/R_{E}}$ is the stellar
equatorial radius projected on the lens plane and normalized to the
Einstein radius $\mathrm{R_{E}}$, $\mathrm{x_{rel}=D_{l}/D_{s}}$ is
the ratio of the lens distance to the source distance from the
observer position,
$\mathrm{u=|\boldsymbol{u}_{cm}-\boldsymbol{\rho}\rho_{\star}|}$ is
the distance of each projected element over the source surface with
respect to the lens position, $\mathrm{\boldsymbol{u}_{cm}}$ is the
vector of the lens position from the source center and
$\mathrm{\varphi}$ is the angle of the normal vector to the
co-center ellipse passing the point $\mathrm{(\rho,\phi)}$ (i.e.
$\mathrm{\boldsymbol{n}}$) with respect to the
$\mathrm{x_{o}}$-axis. In equation (\ref{tsparam}), we have aligned
the signs of two components of polarized Stokes intensities so that
polarization vectors become tangential to the co-center ellipses.
Finally, the polarization degree $\mathrm{P}$ and the angle of
polarization $\mathrm{\theta_{p}}$ as functions of total Stokes
parameters are \cite{Chandrasekhar60}:
\begin{eqnarray}\label{eq2}
P&=&\frac{\sqrt{S_{Q}^{2}+S_{U}^{2}}}{S_{I}},\nonumber\\
\theta_{p}&=&\frac{1}{2}\tan^{-1}{\frac{S_{U}}{S_{Q}}}.
\end{eqnarray}
\begin{figure}
\begin{center}
\includegraphics[angle=0,width=0.5\textwidth,clip=]{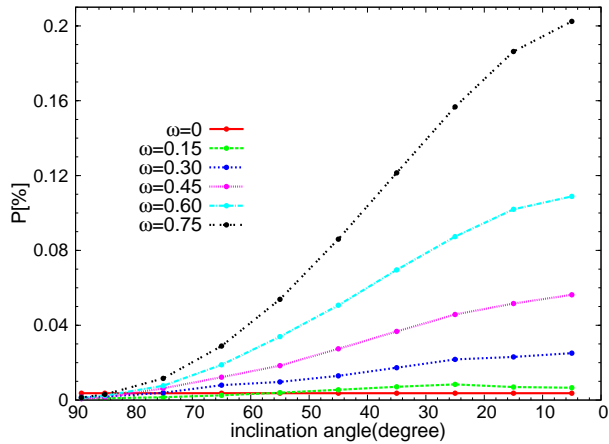}
\caption{The net polarization signal of an elliptical source star
versus the inclination angle of the rotation axis with respect the
sky plane, for different values of $\mathrm{\omega}$ i.e. the ratio
of the stellar angular velocity to the critical velocity. The used
parameters are $M_{\star}=5 M_{\odot}$ and
$\mathrm{R_{p}=3.6~R_{\odot}}$.} \label{fig1}
\end{center}
\end{figure}

If there is no lensing effect, the elliptical source has a net
polarization signal whose magnitude depends on the inclination angle
and its rotational speed. In Figure (\ref{fig1}) we plot the net
polarization signal of an elliptical source versus the inclination
angle for different values of $\mathrm{\omega}$. This plot is the
same as Figure (4) of Harrington \& Collins (1968) representing the
polarization signal of rotating stars. The inclination angle of the
rotational axis in our formalism is the complementary angle to the
one introduced by Harrington \& Collins. If $\mathrm{i=90^{\circ}}$,
the projected shape of the source star on the sky plane is a perfect
circle with the radius equals to $\mathrm{R_{eq}}$. If
$\mathrm{i=0^{\circ}}$, the projected shape will be an ellipse whose
semimajor and minor axes are $\mathrm{R_{eq}}$ and $\mathrm{R_{p}}$.
Therefore, by decreasing the inclination angle the ellipticity of
the source surface projected on the sky plane increases. On the
other hand, faster rotating stars are more oblate. When
$\mathrm{\omega=0}$ the source star is a perfect sphere without any
gravity-darkening effect. In that case, the inclination angle does
not alter the net polarization signal. Noting that the stellar polar
radius is determined using the mass-radius relation while its
equatorial radius depends on $\mathrm{\omega}$ (see equation
\ref{Rstar}). 
Also, the polarization angle of an elliptical source is zero with
respect to its semimajor axis and it does not depend on the
inclination angle and the stellar angular velocity.

According to Figure (\ref{fig1}), the intrinsic polarization signal
produced by an elliptical source is generally less than
$\mathrm{0.2\%}$ and in the case of very fast rotating stars
($\mathrm{\omega>0.6}$) comparable with the polarization signal in
the transit microlensing events for which $\mathrm{P \simeq
0.6-0.7\%}$. Polarization signals of this amplitude can be measured
directly. Therefore, the stellar rotating effects should be
considered in the polarimetric microlensing calculations of fast
rotating stars allowing to characterize the elliptical properties of
the source stars. On the other hand, the intrinsic polarization
signals for slow or moderate rotating stars with
$\mathrm{\omega<0.6}$ that are too small to be measured by
themselves, may be magnified during a microlensing event. In the
next section, we add the lensing effect and study the polarimetry
microlensing of elliptical sources. Our aim is to investigate if
these signals can be detected in polarimetry and photometry
observations of high-magnification microlensing events.
\begin{figure}
\begin{center}
\includegraphics[angle=0,width=8.cm,clip=]{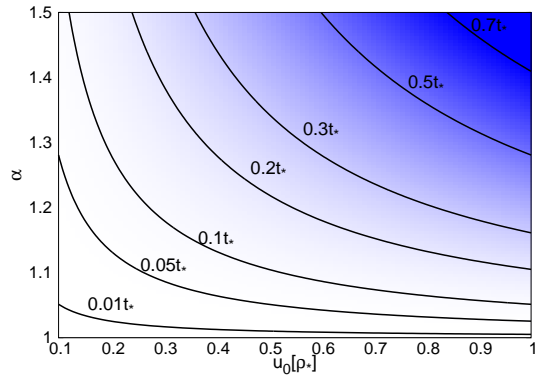}
\caption{The contour lines of the time shift in polarimetry peak
positions of transit microlensing events of elliptical source stars
given by the equation (\ref{times}), in the plane containing
$\mathrm{\alpha}$ (the ratio of the semimajor to semiminor axes of
the projected source surface) and the lens impact parameter
normalized to $\mathrm{\rho_{\star}}$. The contour lines are in the
unit of the time scale of crossing the source radius
$\mathrm{t_{\star}}$.} \label{fig2}
\end{center}
\end{figure}

\section{Characteristics of polarimetric microlensing of elliptical
sources}\label{four} The stellar rotation causes (i) the ellipticity
and (ii) the gravity-darkening effects both of which make
perturbations in the light and polarimetric curves of microlensing
events. Here, we aim to study these perturbations which are
classified in the following subsections.

\subsection{Time shift in the polarimetry peak location}\label{32}
In the case of spherically symmetric (non rotating) source stars,
photometric and polarimetric light-curves in single-lens
microlensing events have symmetric shapes with respect to the time
$\mathrm{t_{0}}$ of the closest approach. In the case of rotating
source stars, the ellipticity effect breaks the symmetry of these
curves around $\mathrm{t_{0}}$.

In particular, (i) the location of the peak(s) in the transit and
bypass polarimetry microlensing curves changes from the points
$\mathrm{u\approx0.96\rho_{\star}}$ and $\mathrm{u=u_{0}}$
respectively, and (ii) the location of the peak in light curves
shifts from the time of the closest approach. But, we can not
realize the shift in the position of the photometric peak from its
true value (i.e. the time of the closest approach) by doing only
photometry observation of microlensing events. While by performing
polarimetry observations of transit microlensing events, we can
discern the time shift in the position of the second peak with
respect to the symmetric (with respect $\mathrm{t_{0}}$) position of
the first peak, since the symmetry in positions of polarimetry peaks
breaks due to the stellar ellipticity. Also in transit microlensing
events the ellipticity can shift the relative minimum location
between two polarimetry peaks with respect to the photometry peak
position. In bypass microlensing events, the time shift between the
position of the polarimetry peak and the position of the photometry
peak helps to distinguish the ellipticity effect. However, in the
bypass polarimetry microlensing events with $\mathrm{u_{0}}$ so
larger than $\mathrm{\rho_{\star}}$ the rotation-induced
perturbations becomes too small, because of the averaging process.

As we shall see in subsection (\ref{33}), gravity-darkening breaks
the symmetry of the source surface brightness and affects on the
peak position of polarimetry and light curves, unless
$\mathrm{i=90^{\circ}}$. However, if the inclination angle is large
enough so that the projected position of stellar pole on the sky
plane (i.e. $\mathrm{\boldsymbol{d}_{R}}$) does not posit at the
stellar limb, we can ignore the gravity-darkening effect for
calculating the time shift in peak positions of polarimetry and
light curves. In that case, by considering only the ellipticity
effect we assess the time shift in the position of the second
polarimetric peak. In transit microlensing events, the lens crosses
the source edges at two moments which are given by:
\begin{eqnarray}
t_{1,2}[t_{E}]= \frac{-u_{0}\sin(2\xi)(\alpha^{2}-1)\pm
2\alpha\sqrt{b^2\cos^{2}\xi+a^2\sin^{2}\xi-u_{0}^{2}}}{2(\cos^{2}\xi+\alpha^{2}\sin^{2}\xi)},
\end{eqnarray}
where $\mathrm{t_{1,2}[t_{E}]}$ are in the unit of the Einstein
crossing time $\mathrm{t_{E}}$, $\mathrm{\mathrm{\xi}}$ is the angle
between the lens trajectory and the source semimajor axis and
$\mathrm{\alpha=a/b}$ the ratio of the source semimajor to minor
axes. Hence, the second peak in polarimetry curves with respect to
the symmetric position of the first peak shifts due to the
ellipticity by:
\begin{eqnarray}\label{times}
\delta_{t}=\frac{u_{0}[\rho_{\star}]\sin(2\xi)(\alpha^2-1)}{(\cos^{2}\xi+\alpha^{2}\sin^{2}\xi)}t_{\star},
\end{eqnarray}
where $\mathrm{u_0[\rho_{\star}]}$ is the lens impact parameter in
units of $\mathrm{\rho_{\star}}$ and
$\mathrm{t_{\star}=\rho_{\star}t_{E}}$ is the time scale for
crossing the source radius. The time shift $\mathrm{\delta_{t}}$
maximizes when $\mathrm{\cos \xi=\alpha/ \sqrt{\alpha^{2}+1}}$. In
Figure (\ref{fig2}), we plot the contour lines of
$\mathrm{\delta_{t}}$ in the plane containing $\mathrm{u_{0}}$ and
$\mathrm{\alpha}$, for values of the lens trajectory angles
$\mathrm{\xi}$ which offer the maximum values of
$\mathrm{\delta_{t}}$. This time shift is less than
$\mathrm{t_{\star}}$ by two or one order of magnitude.

To estimate the magnitude of $\mathrm{t_{\star}}$ and
$\mathrm{\delta_{t}}$, we consider early-type stars as microlensing
sources which most rotate fast. These stars have the effective
temperature in the range of $\mathrm{T_{eff}\in [7500:30000]K}$ and
the stellar polar radius in the range of $\mathrm{R_{p}\in
[1.4:6.6]R_{\odot}}$. It is well known that the abundance of
early-type stars in the Galactic spiral arms is more than that in
the Galactic bulge. Therefore, we consider hypothetical microlensing
events toward the Galactic spiral arm and in the direction of
$\mathrm{l=300^{\circ}}$ and $\mathrm{b=-1^{\circ}}$, i.e. the
Carina-Sagittarius arm. This arm mostly contains young stellar
objects (e.g. Churchwell et al. 2009). By using a Monte Carlo
simulation we estimate the mean lens mass
$\mathrm{M_{l}=0.3M_{\odot}}$ from the Kroupa mass function
\cite{kroupa01,kroupa93} and the average distances
$\mathrm{D_{l}=4.0~kpc}$ and $\mathrm{D_{s}=8.5~kpc}$ of lenses and
sources from the observer by using the angular distribution of stars
in the Galactic bulge and disk. In this way the Einstein radius
results to be $\mathrm{R_{E}=2.3~AU}$. Moreover, by adopting the
synthetic Besan\c{c}on model \cite{besancon} we estimate the source
star radius $\mathrm{\rho_{\star}}$, which is in the range
$\mathrm{[0.001:0.006]}$ (in units of $\mathrm{R_{E}}$). Towards
this Galactic spiral arm the averaged amount of the Einstein
crossing time is longer than that toward the Galactic bulge and is
about $\mathrm{\bar{t}_{E}\simeq97 days}$ \cite{Rahal2009}.
Consequently, for microlensing events toward this Galactic spiral
arm the value of $\mathrm{t_{\star}}$ will be in the range of
$\mathrm{t_{\star}\in[2.3:14.0]}$ hours. Considering a common value
for the time shift $\mathrm{\delta_{t}\sim 0.1 t_{E}}$, it will be
in the range $\mathrm{\delta_t\in[14.0:83.8]~min}$.

To evaluate how many polarimetry data points can potentially be
taken during this time shift $\mathrm{\delta_{t}}$, we assume that
these observations are done by the FOcal Reducer and low dispersion
Spectrograph (FORS2) polarimeter at Very Large Telescope (VLT)
telescope. The necessary exposure time for FORS2 to achieve the
polarimetric accuracy $\mathrm{0.1\%}$ for a magnified star with the
apparent magnitude about $\mathrm{m_{I}=14.5~mag}$ is about
$\mathrm{8~sec}$
\footnote{http://www.eso.org/observing/etc/bin/gen}. In addition to
the exposure time, there are two extra waste times due to the
retarder waveplate rotation and CCD readout. Indeed, to accurately
determine the polarization signal, the source flux should be
measured in $16$ directions from $\mathrm{0^{\circ}}$ to
$\mathrm{337.5^{\circ}}$, in $\mathrm{22.5^{\circ}}$ steps. The
signal to noise ratio ($\mathrm{S/N}$) of the accumulated flux from
total taken exposure time in all retarder waveplate positions should
reach to $\mathrm{1000}$ to give up the polarimetry accuracy
$\mathrm{0.1\%}$ \cite{Ejeta2012}. Rotating the retarder waveplate
of FORS2 takes some time about $\mathrm{\sim1~min}$. On the other
hand, the FORS2 CCD will be saturated after $\mathrm{2~sec}$
exposure time from a bright star with $\mathrm{m_{I}=14.5~mag}$.
Therefore, the CCD detector should be read $\mathrm{4}$ times for
taking every polarimetry data point with the polarimetry accuracy
$\mathrm{0.1\%}$. The FORS2 CCD readout takes time
$\mathrm{30~sec}$. Accordingly, we should add to the exposure time
about $\mathrm{18~min}$ as overhead time due to the retarder
waveplate rotation and CCD readout. This overhead time does not
depend on the magnification factor and the source brightness and is
constant for taking each data point by FORS2 with the highest
polarimetry accuracy. Thus, the total observational time for each
polarimetry data point of a magnified source star with
$\mathrm{m_{I}=14.5~mag}$ is about $\mathrm{T_{obs}=18.13~min}$.

Consequently, If FORS2 uninterruptedly observers a transit
microlensing event of an early-type star, it will averagely take
$\mathrm{1-5}$ data points during this time shift, where we estimate
the number of possible data points by the factor
$\mathrm{\delta_t/T_{obs}}$. This number is not sufficient to
correctly realize $\mathrm{\delta_{t}}$. However, the factor
$\mathrm{\delta_t/T_{obs}}$ increases in some specific microlensing
events, e.g. when (i) the Einstein crossing time is very long, (ii)
the lens crosses the source surface with large impact parameters
(see Figure \ref{fig2}) and (iii) the projected radius of the source
star normalized to the Einstein radius is large which happens e.g.
when the lens and source stars are so close to each other i.e.
$\mathrm{x_{rel}\sim1}$. Otherwise, the possibility of discerning
this time shift is almost out of the present technology (FORS2
polarimeter at VLT telescope). However, this time shift can likely
be realized by high-quality instruments of the next generation.

\begin{figure*}
\centering
\subfigure[] {
\includegraphics[angle=0,width=8.cm,clip=]{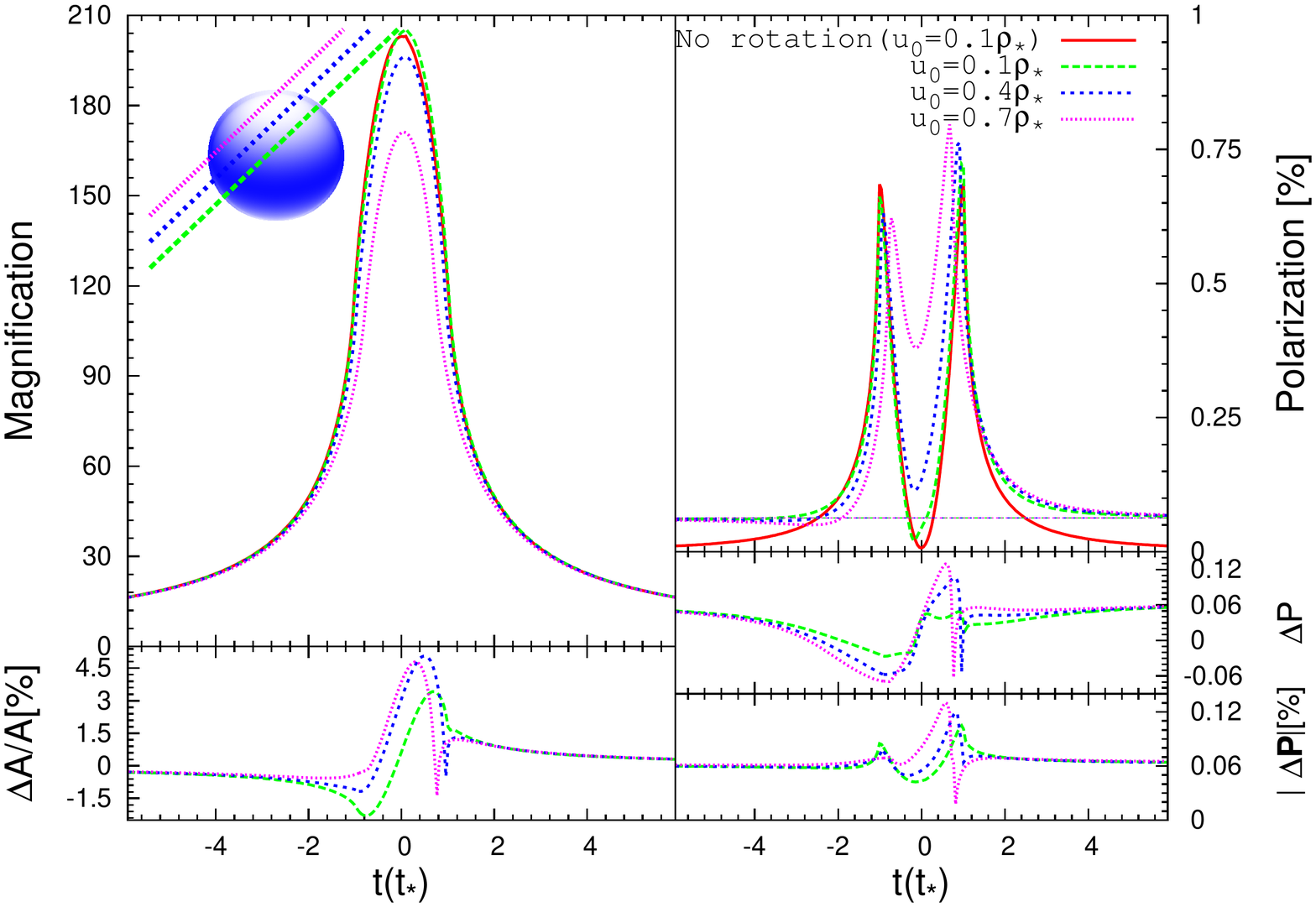}
\label{fig4a} } \subfigure[] {
\includegraphics[angle=0,width=8.cm,clip=]{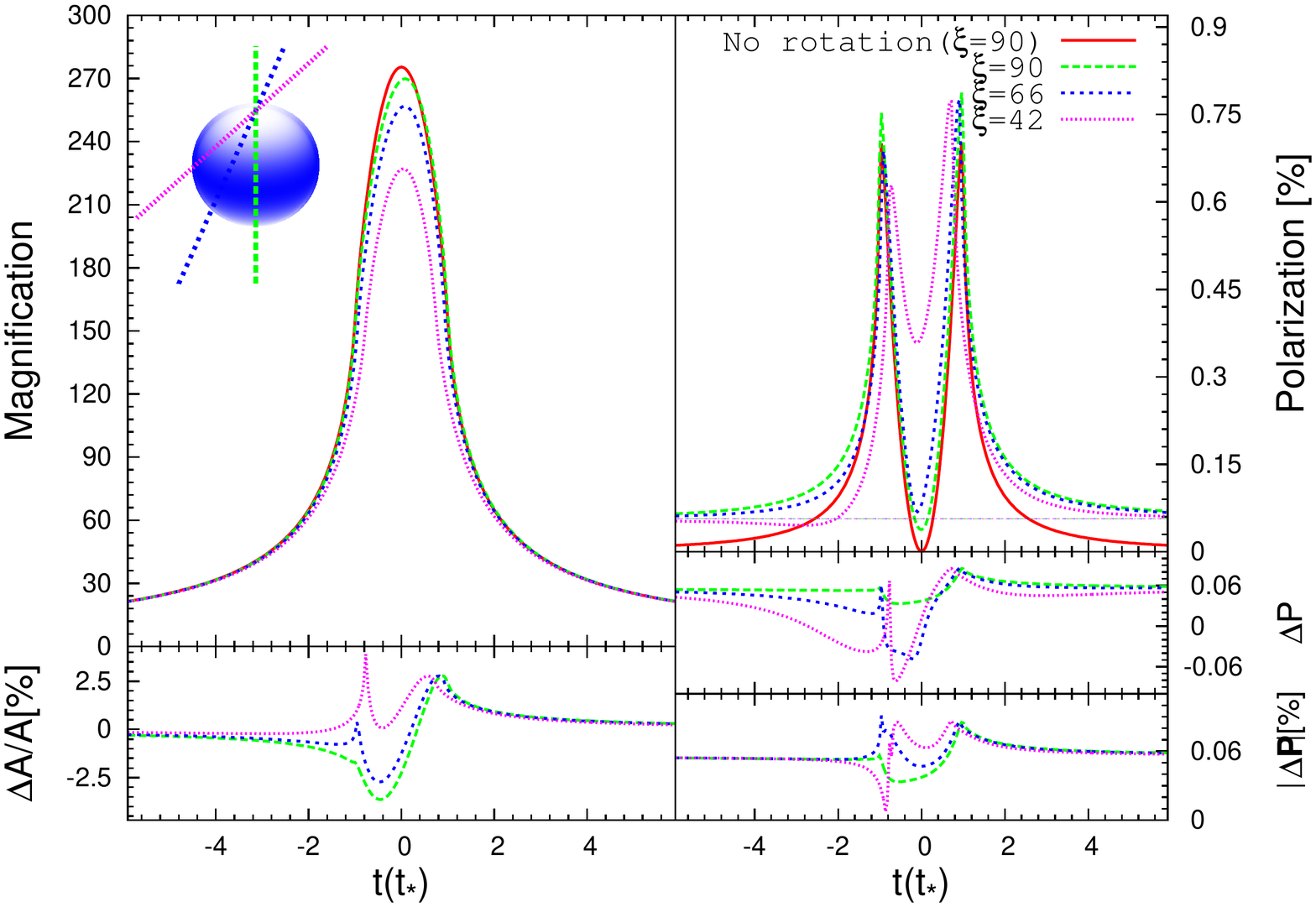}
\label{fig4b}} \subfigure[] {
\includegraphics[angle=0,width=8.cm,clip=]{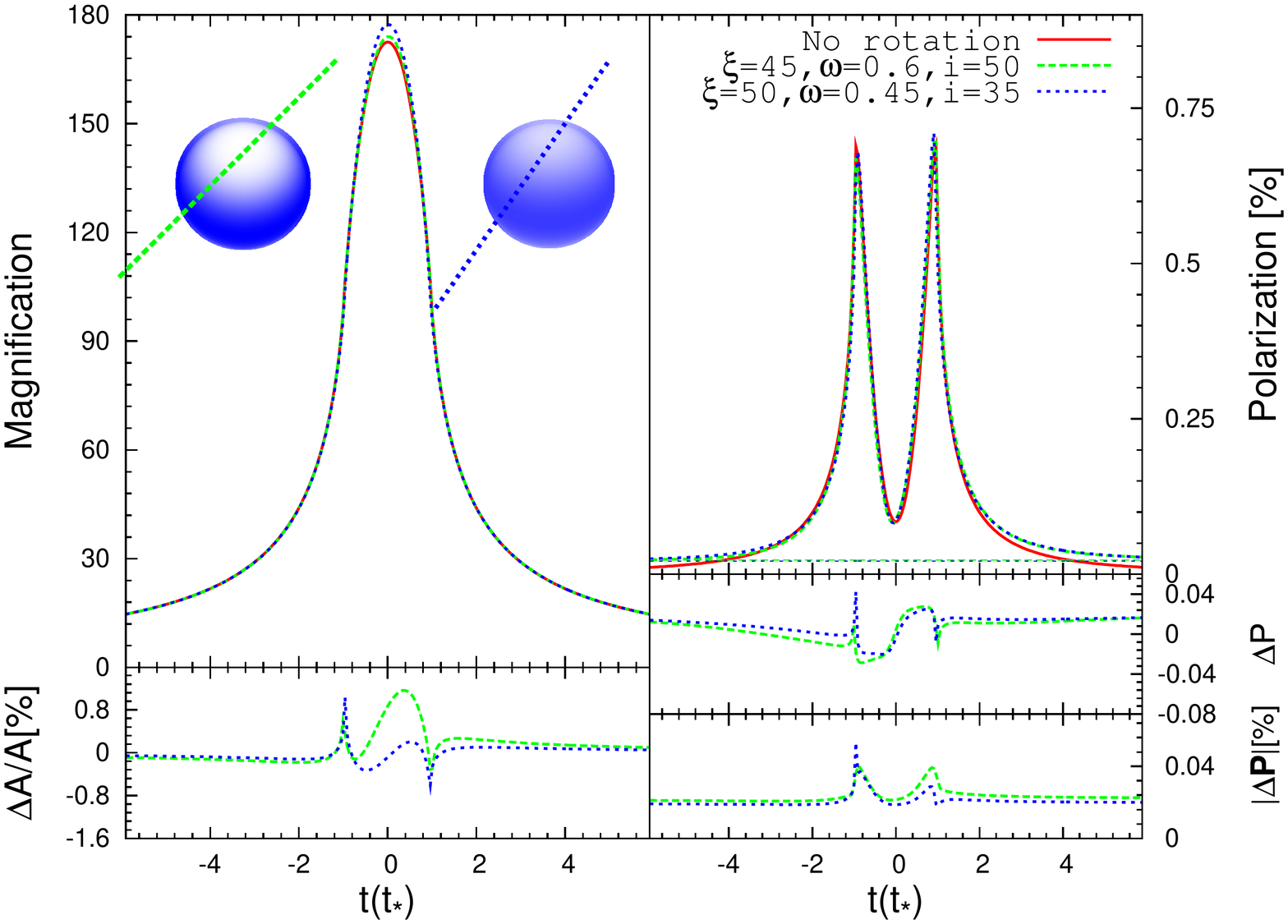}
\label{fig4c} } \subfigure[] {
\includegraphics[angle=0,width=8.cm,clip=]{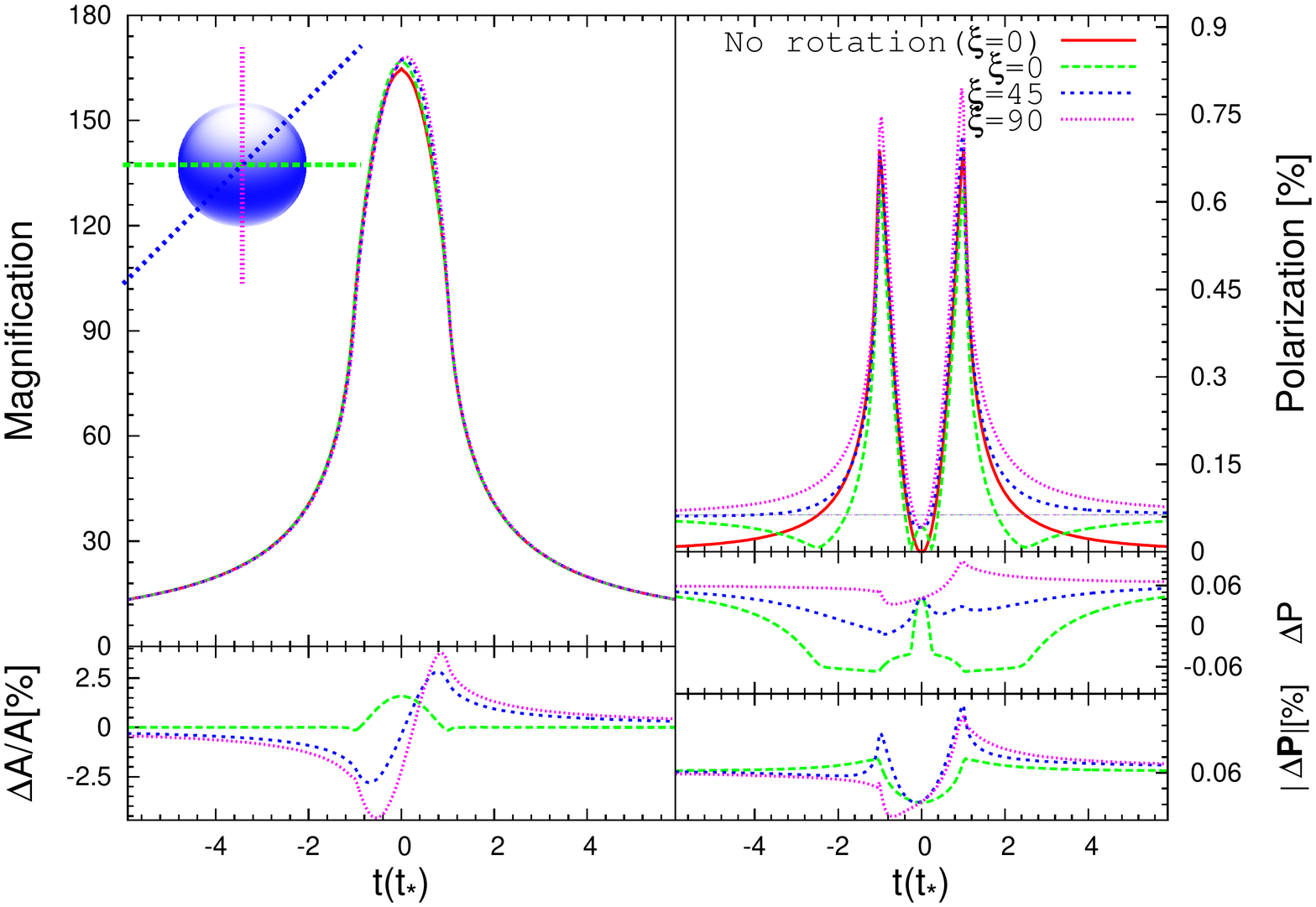}
\label{fig4d} }\caption{Example polarimetric microlensing events of
elliptical source stars. In every subfigure, the light and
polarimetric curves are shown in left and right panels. The source
surface projected on the lens plane and the lens trajectory are
shown with the inset in the left-hand panel. The map over the source
surface represents stellar surface brightness. The simple models
without the stellar rotation are plotted by red solid lines. The
thinner straight lines in the right panel represent the intrinsic
polarization signal of the elliptical source stars. The photometric
and polarimetric residuals with respect to simple models are plotted
in bottom panels. The parameters used to make these figures can be
found in Table (\ref{tab1}). We also set
$\mathrm{M_{l}=0.3~M_{\odot}}$, $\mathrm{D_{l}=6.5~kpc}$,
$\mathrm{D_{s}=8~kpc}$ and the limb-darkening coefficients over the
source surface $\mathrm{c_{1}=0.64}$ and
$\mathrm{c_{2}=0.032}$.}\label{fig4t}
\end{figure*}

If the time shift $\mathrm{\delta_{t}}$ is measured from the
polarimetric observations of microlensing events, we can estimate
the parameter
$\mathrm{\alpha=R_{eq}/\sqrt{R_{eq}^{2}\sin^{2}(i)+R_{p}^{2}\cos^{2}(i)}}$
which is a degenerate function of the inclination angle and the
intrinsic ellipticity of the source star, assuming the lens impact
parameter and $\mathrm{\xi}$ are carefully measured from photometry
observations. The parameter $\mathrm{\alpha}$ shows the ellipticity
of the source surface projected on the sky plane. Noting that the
uncertainties in the parameters $\mathrm{u_{0}}$, $\mathrm{\xi}$ and
$\mathrm{\delta_{t}}$ cause an uncertainty in the parameter
$\mathrm{\alpha}$. The uncertainty in $\mathrm{\delta_{t}}$ is due
to time intervals between consecutive polarimetry data points
hypothetically taken during this time shift and their uncertainties.

\begin{deluxetable}{ccccccc}
\tablecolumns{7} 
\tablewidth{0.45\textwidth} \tabletypesize \footnotesize
\tablecaption{The parameters used to make microlensing events shown
in Figures \ref{fig4a}, \ref{fig4b}, \ref{fig4c}, \ref{fig4d} and
(\ref{fig7}).} \tablehead{ \colhead{} &
\colhead{$\mathrm{M_{\star}(M_{\odot})}$} &
\colhead{$\mathrm{R_{p}(R_{\odot})}$} & \colhead{$\mathrm{\omega}$}
& \colhead{$\mathrm{i^{\circ}}$}&
\colhead{$\mathrm{u_{0}(\rho_{\star})}$}&
\colhead{$\mathrm{\xi^{\circ}}$}} \startdata
$3(a)$ & $7.0$ & $4.7$ & $0.55$  & $30.0$ & $--$ & $43.9$ \\
\\ 
$3(b)$ & $5.0$ & $3.6$ & $0.5$ & $25.0$ & $0.9\cos(\xi)$ & $--$ \\
\\ 
$3(c)$ & $8.0$ & $5.3$ & $--$  & $--$ & $0.35$ & $--$  \\
\\ 
$3(d)$ & $9.0$ & $5.8$ & $0.55$ & $30.0$ & $0.0$ & $--$ \\
\\ 
$(7)$ & $3.8$ & $2.9$ & $0.50$ & $15.0$ & $0.43$  & $42.0$  \\
\enddata
\tablecomments{The columns contains (i) the figure number, (ii) the
mass of the source star $\mathrm{M_{\star}(M_{\odot})}$, (iii) the
polar source radius $\mathrm{R_{p}(R_{\odot})}$, (iv) the angular
speed of the source normalized to the break-up velocity
$\mathrm{\omega}$, (v) the inclination angle of the stellar
rotational axis with respect to the sky plane $\mathrm{i^{\circ}}$,
(vi) the impact parameter of the lens trajectory with respect to the
source center normalized to $\mathrm{\rho_{\star}}$
$\mathrm{u_{0}(\rho_{\star})}$ and (vii) the angle of the lens
trajectory with respect to the source semimajor axis
$\mathrm{\xi^{\circ}}$ respectively. We also set the lens mass
$\mathrm{M_{l}=0.3~M_{\odot}}$, the lens and source distances from
the observer $\mathrm{D_{l}=6.5~kpc}$ and $\mathrm{D_{s}=8.0~kpc}$
for four first figures and $\mathrm{M_{l}=0.7~M_{\odot}}$,
$\mathrm{D_{l}=4.1~kpc}$ and $\mathrm{D_{s}=8.2~kpc}$ for the last
one. The limb-darkening coefficients are fixed at
$\mathrm{c_{1}=0.64}$ and $\mathrm{c_{2}=0.032}$.}\label{tab1}
\end{deluxetable}
In Figure \ref{fig4a} we plot the photometric and polarimetric
light-curves of a microlensing event of an elliptical source for
three different values of the lens impact parameter. In this Figure,
the light and polarimetric curves are shown in the left and right
panels respectively. The projected surface of the source star on the
lens plane and lens trajectories are shown with an inset in the
left-hand panel. The map over the source surface represents the
surface brightness considering the gravity-darkening effect. The
simple models without stellar rotation are shown by red solid lines.
The thinner straight lines in right panel represent the intrinsic
polarization signal of the elliptical source. The photometric and
polarimetric residuals with respect to the simple models are plotted
in bottom panels. The top polarimetric residual is the residual in
the polarization degree $\mathrm{\Delta P=P'-P}$ and the bottom one
is the absolute value of the residual in the polarization vector
$\mathrm{|\boldsymbol{\Delta
P}|=\sqrt{P'^{2}+P^{2}-2P'P\cos2(\theta'_{p}-\theta_{p})}}$, where
the prime symbol refers to the related quantity considering the
stellar rotation. The parameters used to make this figure can be
found in Table (\ref{tab1}). We also set
$\mathrm{M_{l}=0.3~M_{\odot}}$, $\mathrm{D_{l}=6.5~kpc}$,
$\mathrm{D_{s}=8~kpc}$, $\mathrm{c_{1}=0.64}$ and
$\mathrm{c_{2}=0.032}$. For each microlensing event, the radius of
the spherical source of the simple model is equal to
$\mathrm{\rho_{\star,s}=\sqrt{u_{0}^{2}+t^{2}_{1}}}$, i.e. the
radius of the elliptical source where the lens is entering the
source surface. Accordingly, the first peaks of polarimetry curves
are coincided to the first peaks of simple models. The sharp peaks
in polarimetry and photometry residuals while the lens is leaving
the source surface represent the mentioned time shift and are owing
to the ellipticity shape of the source surface which increases with
enhancing the lens impact parameter.

If $\mathrm{u_{0}>\rho_{\star}}$ and for large inclination angles,
the peak position of microlensing light curves of elliptical source
stars does not significantly shift from the time of the closest
approach, since the maximum value of the Stokes intensity
$\mathrm{I_{I}}$ takes place in the stellar center (see equation
\ref{II}) and the ellipticity affects on stellar limb points. On the
contrary, in these bypass microlensing events the polarimetry peak
positions vary from the time of the closest approach due to the
ellipticity of the source surface, since the maximum value of the
polarized Stokes intensity $\mathrm{I_{p}}$ occurs at the stellar
limb (see equation \ref{II}) and the stellar ellipticity also
affects on these points. We expect that the peak position in
polarimetric curves happens where the distance between the lens and
the source edge minimizes. Accordingly, the time shift between
polarimetric and photometric peaks is given by:
\begin{eqnarray}
\delta_{t}=(\alpha-\frac{1}{\alpha})\frac{\cos\xi}{\sqrt{\alpha^{2}+\cot^{2}\xi}}~t_{\star}.
\end{eqnarray}
The maximum value of this time shift happens when
$\cot\xi=\sqrt{\alpha}$ which is equal to $\mathrm{0.33t_{\star}}$
when $\mathrm{\alpha=1.5}$. However, the gravity-darkening affects
on this time shift when the inclination angle is small.
\begin{figure}
\begin{center}
\includegraphics[angle=0,width=8.cm,clip=]{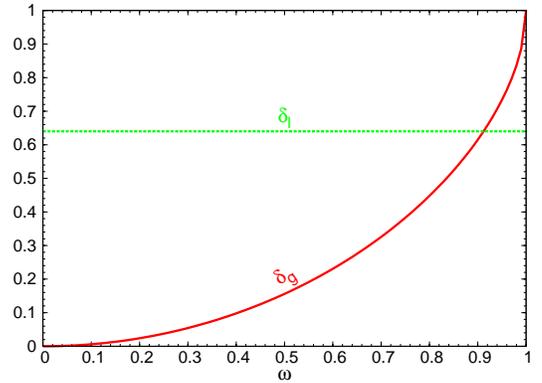}
\caption{The maximum amounts of the relative deviations in the
stellar surface brightness due to the gravity-darkening effect (red
solid line) and owing to the limb-darkening effect (green dashed
line) versus $\mathrm{\omega}$.} \label{fig3}
\end{center}
\end{figure}

\subsection{Asymmetric perturbations owing to the gravity-darkening
effect}\label{33} Gravity-darkening breaks the circular symmetry of
the source surface brightness. However, gravity-darkening does not
break the axial symmetry of the projected source surface with
respect to its semiminor axis ($\mathrm{y_{o}}$-axis). This effect
causes asymmetric perturbations in photometric and polarimetric
light-curves of microlensing events, unless $\mathrm{\xi=0^{\circ}}$
or $\mathrm{i=90^{\circ}}$. The gravity-darkening effect can be
evaluated by the relative maximum deviation in the source surface
brightness i.e. $\mathrm{\delta_{g}}$ which is given by:
\begin{eqnarray}
\delta_{g}=|\frac{g_{eff}(\Omega,0)-g_{eff}(\Omega,90^{\circ})}{g_{eff}(\Omega,0)}|=
1-\frac{R_{p}^{2}}{R_{eq}^{2}}+\omega^{2}(\frac{2}{3})^{3}\frac{R_{eq}}{R_{p}},
\end{eqnarray}
which depends only on $\mathrm{\omega}$ i.e. the stellar angular
velocity normalized to the break-up velocity. On the other hand, the
relative maximum deviation in the stellar surface brightness due to
the limb-darkening effect is $\mathrm{\delta_{l}=c_{1}(=0.64)}$. To
compare the gravity-darkening and limb-darkening effects, we plot
these relative maximum deviations in Figure (\ref{fig3}), in which
the red solid line represents $\mathrm{\delta_{g}}$ and the
horizontal green dashed line shows $\mathrm{\delta_{l}}$. Thus, the
gravity-darkening effect generally is so smaller than the
limb-darkening effect. It mostly makes small perturbations in
microlensing curves, unless the lens crosses the projected position
of the stellar rotation pole i.e. $\mathrm{\boldsymbol{d}_{R}}$. In
these cases in which the lens impact parameters equal to
$\mathrm{u_{0}=R_{p}\cos(i)\cos(\xi)}$, the gravity-darkening effect
can make detectable perturbations. The maximum deviation in
microlensing light curves due to the gravity-darkening effect
happens when $\mathrm{\xi=90^{\circ}}$, since the maximum Stokes
intensity $\mathrm{I_{I}}$ which occurs at the source center (due to
limb-darkening effect) is significantly perturbed by the
gravity-darkening effect. The maximum polarimetric perturbation
takes plane when $\mathrm{\tan(\xi)= R_{p}\cos(i)/a}$, because in
that case the lens trajectory crosses the stellar edge points which
have the maximum value of $\mathrm{\delta_{g}}$ while these points
own the largest polarized Stokes intensity. These asymmetric
perturbations in polarimetry and photometry curves of microlensing
events can be identified by comparing the left and right sides of
these curves.

We show the maximum asymmetric perturbations in the light and
polarimetry curves of a microlensing event due to the
gravity-darkening effect in Figure \ref{fig4b}. The
characterizations of this plot are the same as those of Figure
\ref{fig4a}. We align the source trajectory so that it crosses the
point $\mathrm{\boldsymbol{d}_{R}}$ for three different values of
$\mathrm{\xi}$. When $\mathrm{\xi=90^{\circ}}$ the photometric
residual due to the gravity-darkening effect maximizes. When
$\mathrm{\xi=42^{\circ}}$, the polarimetry residual becomes maximum.
Asymmetry in microlensing curves due to the gravity-darkening effect
is obvious, because the polarimetry and photometry residuals are not
symmetric with respect to the time of the maximum magnification.
Note that the sharp peaks when the lens is entering the source
surface are due to the ellipticity effect and the mentioned time
shift. However, the polarimetric deviations due to the
gravity-darkening in this figure are less than the polarimetry
precision of FORS2 which means higher quality polarimeters can
realize these rotation-induced perturbations.

There is a problem in microlensing observations which is degeneracy.
Even if microlensing observers correctly discern the type of the
anomaly, all parameters of this anomaly can not be uniquely derived
from the observed light or polarimetric curves. In polarimetry or
photometry microlensing events of elliptical source stars this
degeneracy exits. The intrinsic polarization signal of an elliptical
source is a degenerate function of the inclination angle and the
stellar angular velocity (see Figure \ref{fig1}). This degeneracy
can not be resolved in microlensing observations. Figure \ref{fig4c}
represents two different microlensing events of elliptical sources
with different parameters, but the same polarimetry and photometry
curves. These two microlensing events are degenerate. However, the
microlensing degeneracy can even exit between different models with
different kinds of anomalies. For example the microlensing curves
and the intrinsic polarization signal of a rotating source star can
be the same as those of two close binary source stars. This point is
not studied in this work.

\begin{figure}
\begin{center}
\includegraphics[angle=0,width=8.cm,clip=]{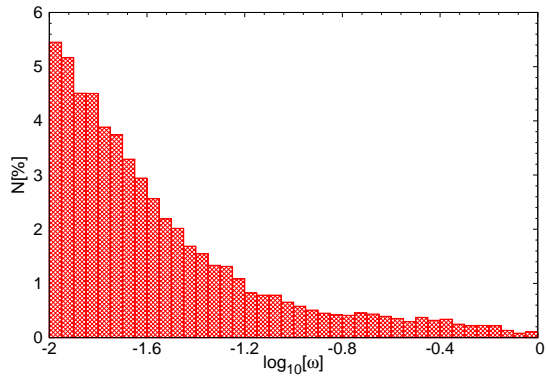}
\caption{Histogram of $\mathrm{\omega}$ in the logarithmic scale for
the main-sequence stars observed by the Kepler satellite and their
rotation periods were estimated by Rienhold et al. (2015).}
\label{fig5}
\end{center}
\end{figure}
The polarization angle of an elliptical source (in our formalism) is
zero with respect to its semimajor axis. The polarization angle of a
lensed source star is $\mathrm{90^{\circ}}$ with respect to the
connection line of the source center and the lens position (see e.g.
Sajadian \& Rahvar 2015). Hence, when the lens is entering the
source surface with $\mathrm{\xi=0^{\circ}}$ these polarization
vectors are normal and in some time eliminate each other, so that
the total polarization signal tends to zero at that time. Whereas,
when $\mathrm{\xi=90^{\circ}}$ these two polarization vectors are
parallel and always magnify each other, so that the total
polarization signal is ascending while the lens is entering the
source surface. These points are shown in Figure \ref{fig4d}.
Detecting these features in the polarimetry curves of microlensing
events helps to discern the angle between the lens trajectory and
the semimajor axis of elliptical source which breaks the
microlensing degeneracy. In this figure we set $\mathrm{u_{0}=0}$,
so peak positions of polarimetry and photometry curves have no time
shift. However, there are some asymmetric perturbations due to the
gravity-darkening effect. The photometric perturbation due to the
ellipticity maximizes when $\mathrm{\xi=90^{\circ}}$.

According to the different panels of Figure (\ref{fig4t}), it seems
that most rotation-induced perturbations in polarimetric curves of
microlensing events are less than the FORS2 accuracy i.e.
$\mathrm{0.1\%}$. Hence, detecting the rotation-induced
perturbations in the polarimetry microlensing curves can probably be
done by the next-generation polarimeters with higher precisions than
that of FORS2.

\begin{figure}
\begin{center}
\includegraphics[angle=0,width=8.cm,clip=]{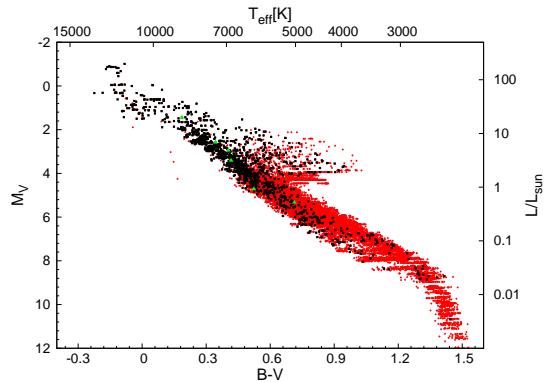}
\caption{The Color-Magnitude diagram of the Kepler stars whose
stellar rotational periods have been specified by Rienhold et al.
(2013,2015) (red points). The stars with $\mathrm{\omega}$ larger
than $\mathrm{0.1}$ are indicated by black stars and the stars have
the intrinsic polarization signals larger than $\mathrm{0.2\%}$ are
represented by green triangles.} \label{fig6}
\end{center}
\end{figure}

\section{Observational remarks}\label{five}
in the previous section we studied some aspects of polarimetry and
photometry microlensing events of elliptical sources. In this
section, we first investigate how per cent of the magnetically
active stars observed by the Kepler satellite rotate fast and have
the considerable values of $\mathrm{\omega}$. The rotational periods
of these stars were evaluated by their light curves by some groups
(e.g. Reinhold et al. 2013, 2015 \& McQuillan et al. 2014). Then, we
study whether the polarimetry observations by FORS2 of
high-magnification microlensing events of fast rotating stars
towards the Galactic spiral arms can give information on
rotation-induced perturbations.

\subsection{Statistic of fast rotating stars based on the Kepler data}
The accurate statistics of fast rotating stars can not be fully
determined owing to several limitations in observational methods.
For example, the photometry method for measuring stellar rotational
periods is sensitive to magnetically active stars with stellar
spots. Also, the interferometry method evaluates the stellar
oblateness of just nearby stars.

The Kepler satellite has provided stellar light curves of a very
large sample of stars for more than four years by doing
high-resolution uninterrupted photometry observations
\cite{borucki10,koch10}. These data were analyzed to derive the
stellar rotation periods and differential rotation effects by
several authors, e.g. Reinhold et al. (2013,2015) analyzed a large
sample of the Kepler stars to determine their rotation periods using
different approaches based on the Lomb-Scargle periodogram. They
first selected magnetically active stars which exhibit stellar spots
at their surface, i.e. often main-sequence stars with
$\mathrm{\log_{10}(g)>3.5}$. They noticed that $\mathrm{24.6\%}$ of
stars in their sample were active. 
Using the stellar rotation periods given by Reinhold et al., the
angular velocities of these active stars are inferred. We also
estimate their critical velocities $\mathrm{\Omega_{crit}}$ and as a
result the ratio of the stellar angular velocity to the critical
velocity $\mathrm{\omega}$, according to the mass and radius of
these stars given by Huber et al. (2014). The distribution of
$\mathrm{\log_{10}(\omega)}$ for this sample of stars is plotted in
Figure (\ref{fig5}). About $\mathrm{3.7}$ and $\mathrm{6.6}$ per
cent of these stars have $\mathrm{\omega}$ larger than
$\mathrm{0.2}$ and $\mathrm{0.1}$ respectively. This sample just
contains active main-sequence stars with
$\mathrm{\log_{10}(g)>3.5}$, but not all unbiased stars.

\begin{figure}
\begin{center}
\includegraphics[angle=0,width=8.cm,clip=]{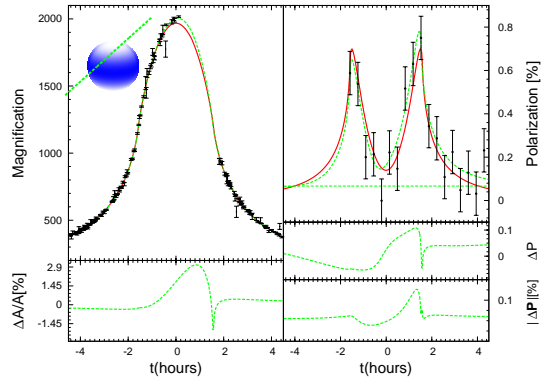}
\caption{An example high-magnification polarimetry microlensing
event and its simulated data points hypothetically taken by surveys
and follow-up telescopes (over its light curve) and FORS2
polarimeter (over its polarimetry curve).} \label{fig7}
\end{center}
\end{figure}
The Color-Magnitude (CM) diagram of these Kepler stars is shown in
Figure (\ref{fig6}) (red points). In this figure, the stars with
$\omega$ larger than $0.1$ are shown with black stars. Most of fast
rotating stars are early-type and hot stars. The green triangles
represent the stars with the intrinsic polarization signals larger
than $0.2\%$, i.e. their polarization signals are measurable by
FORS2 even without the lensing effect. For stars of this sample, we
estimate their absolute magnitude using a synthetic CM diagram. We
first generate a big ensemble of stars using the isochrones of
Padova \cite{Marigo2008} and according to the strategy explained in
section (3) of Sajadian \& Rahvar (2012). We compare the
temperature-luminosity diagram of stars in this sample with the
related diagram of the generated synthetic sample of stars. For each
star in our sample, we pick the characteristics of the most similar
synthetic star in the generated ensemble.

In the next subsection, we simulate high-magnification microlensing
of the stars specified by the black stars in Figure (\ref{fig6}) to
study if rotation-induced perturbations are discernable through
hypothetical polarimetry and photometry observations of these
events.

\subsection{Monte Carlo simulation}
It is well known that massive and hot stars rotate very fast,
whereas most main-sequence or red giant stars have small or moderate
rotational speeds (e.g. Bouvier 2013). The Galactic bulge often
contains old and cold stars while the Galactic disk stars have a
wide range of ages (e.g. Ortolani et al. 1995, Russeil 2003).
Indeed, our galaxy seems to have two spiral arms containing old
stars and four spiral arms including gas and young stars
\cite{urquhart2014}. Although, the microlensing optical depth toward
the Galactic spiral arms is less than that toward the Galactic bulge
by one order of magnitude \cite{Rahal2009}, the mean duration of
microlensing events in these directions is $\mathrm{\sim60~days}$
\cite{Rahal2009} which is longer than that toward the Galactic
bulge, i.e. $\mathrm{\sim 27~days}$ \cite{ogle3}.Thus, the time
scale of crossing the source radius by the lens, i.e.
$\mathrm{t_{\star}}$ toward the spiral arms is about twice that
toward the Galactic bulge. We note that $\mathrm{t_{\star}}$ is the
polarimetry time scale during a microlensing event. On the whole,
the Galactic spiral arms are more suitable to be probed for finding
stellar rotation effects. Hence, we simulate high-magnification
microlensing events of those source stars whose rotational
properties studied by Reinhold et al. toward the Galactic spiral
arms. We choose the Carina$\mathrm{-}$Sagittarius arm and in the
direction $\mathrm{l=300^{\circ}}$ and $\mathrm{b=-1^{\circ}}$.
Indeed, we assume that the stellar local population probed by the
Kepler satellite is representative of the galactic population. This
hypothesis is most probably justified for stars in the Galactic
disk.

We have two criteria for selecting these stars as source stars: (a)
the stars with $\mathrm{\omega>0.1}$ and (b) those brighter than
$\mathrm{21~mag}$ in I-band after being located at that Galactic
arm. These criteria decrease the number of possible source stars to
$\mathrm{773}$. We finally investigate the possibility of discerning
polarimetry and photometry rotation-induced perturbations in these
high-magnification microlensing events.
\begin{figure*}
\centering 
\includegraphics[angle=0,width=8.cm,clip=]{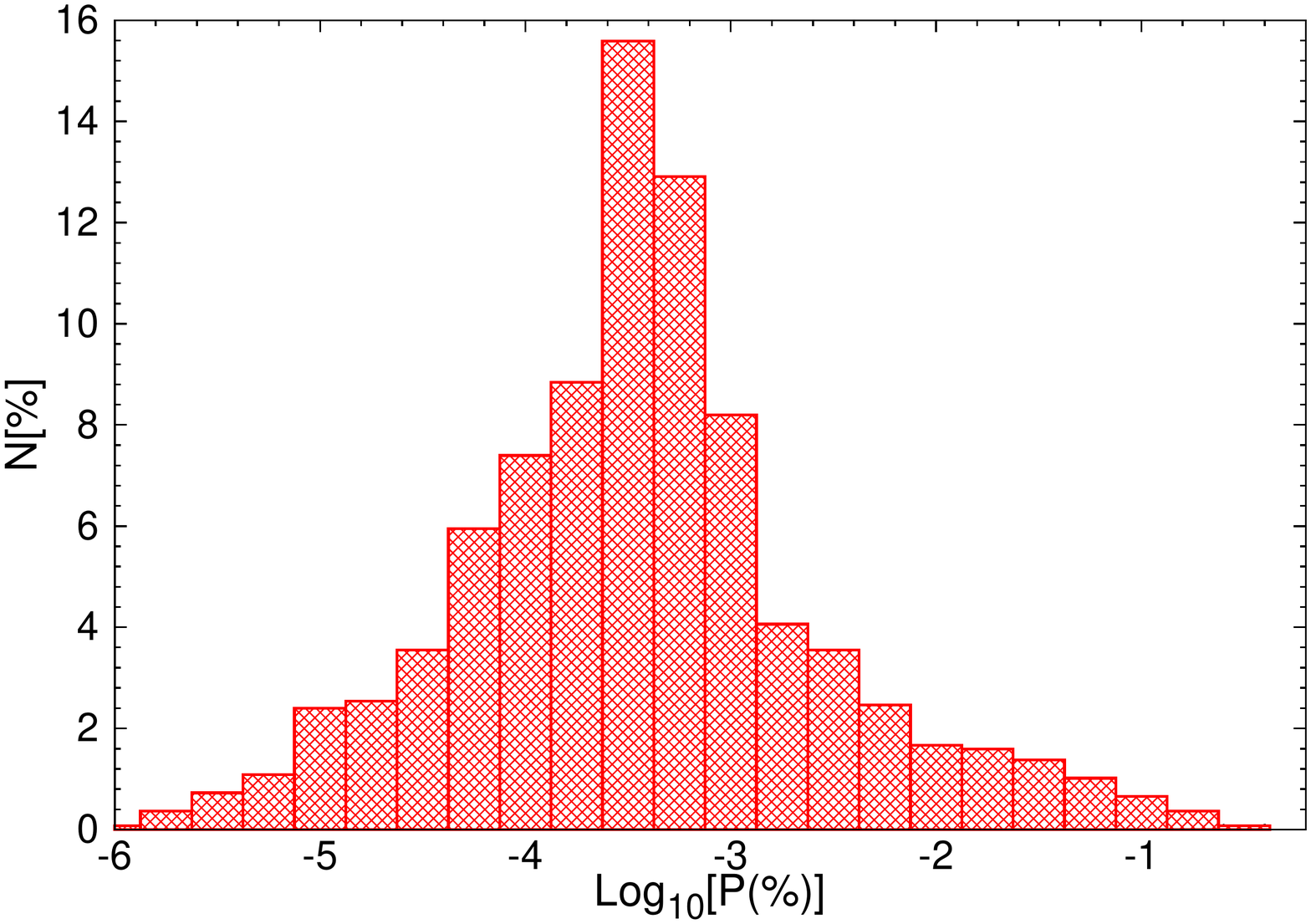}
\includegraphics[angle=0,width=8.cm,clip=]{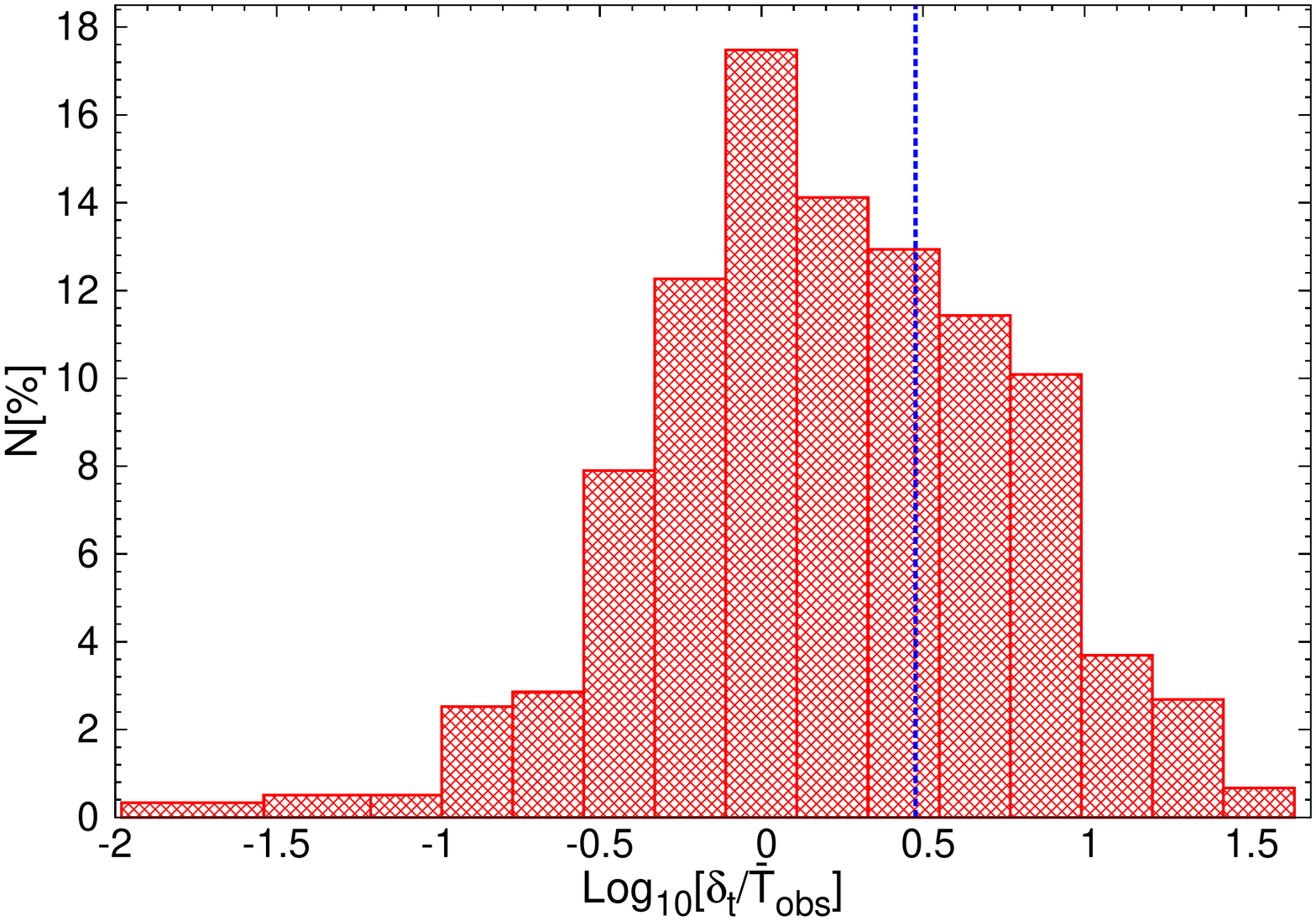}
\caption{The results of the Monte Carlo simulation: Left panel: the
histogram of the intrinsic polarization signals of source stars.
Right panel: the histogram of the time shifts induced by the stellar
ellipticity in the polarimetry peak positions $\mathrm{\delta_{t}}$
normalized to the averaged observational time for taking each
polarimetry data point $\mathrm{\bar{T}_{obs}}$. This quantity shows
the possible number of polarimetric data points during the
rotation-induced time shifts. The blue dashed line shows the amount
$\mathrm{\delta_{t}/\bar{T}_{obs}=3}$. Both of them are plotted in
the logarithmic scale.} \label{fig8}
\end{figure*}

The generic procedure for a Monte Carlo simulation as well as the
used distribution functions to determine the mass of lenses, the
velocities of both sources and lenses and the distribution of matter
in the Galaxy to determine the lens and source positions toward the
Galactic arm were described in our previous works
\cite{sajadian12,sajadian15,Marc2016} and we do not repeat them
here. Also, we use the Galactic extinction model in three dimensions
developed by Marshall et al. (2006). We consider only
high-magnification events with the lens impact parameter less than
the threshold value of $\mathrm{0.001}$. The inclination angle of
the rotational axis for each source star is uniformly chosen in the
range of $\mathrm{[0,90^{\circ}]}$.

We assume these events are observed by FORS2 which reaches the
highest polarimetry precision $\mathrm{\sigma_{p}=0.1\%}$. The
necessary S/N to achieve this precision is
$\mathrm{1/\sigma_{p}=1000}$ for the imaging polarimetry mode
(IPOL). We generate synthetic data points hypothetically taken by
FORS2 over every polarimetry curve. In this regard, we calculate the
necessary exposure time to achieve the highest polarimetric
accuracy. The definition of $\mathrm{S/N}$ can be found in Sajadian
(2015b). In addition to the exposure time, there are two extra waste
times due to the retarder waveplate rotation and CCD readout. This
overhead time lasts about $\mathrm{18~min}$ for each polarimetry
data point by FORS2. In this regard, all details are explained in
subsection (\ref{32}). The start time of observation by FORS2 is
chosen randomly in the range of $\mathrm{[-3.0:3.0]t_{\star}}$ and
after about $\mathrm{6~hours}$, the observation is interrupted until
the next night (after $\mathrm{18~hours}$). For all simulated
events, we set $\mathrm{t_{0}=0}$.

For generating photometry data points, we use the sampling and the
photometric uncertainties (i.e. $\mathrm{\sigma_{a}}$) taken by some
archived high-magnification events around the peak of their light
curves. We use the high-magnification microlensing events given by
Choi et al. (2012). The simulated (photometry and polarimetry) data
points are shifted with respect to the model light and polarimetric
curves according to their photometric and polarimetric uncertainties
by Gaussian functions. One of simulated light and polarimetry curves
of high-magnification microlensing events and its synthetic data
points are shown in Figure (\ref{fig7}). Its parameters can be found
in Table (\ref{tab1}). Also we put $\mathrm{t_{E}=70~days}$ and
$\mathrm{\rho_{\star}=0.001}$. Then, we investigate the simulated
photometry and polarimetry curves to verify if the rotation-induced
perturbations of source stars are distinguishable. There are three
tests in this regard:

(a) If the intrinsic polarization signal of a rotating source star
is larger than $0.2\%$, twice the polarimetric accuracy of the FORS2
polarimeter, the polarization signal can be measured even without
the lensing effect. The distribution of the intrinsic polarization
signals of source stars is shown in the left panel of Figure
(\ref{fig8}). Six source stars have the net polarization signals
more than $0.2\%$. These stars are indicated in Figure (\ref{fig6})
by green triangles.

(b) As discussed in the previous section, the stellar rotation
breaks the symmetry of polarimetry and photometry microlensing
curves with respect to $\mathrm{t_{0}}$. This anomaly shifts (i) the
position of the second polarimetry peak with respect to the
symmetric position of the first polarimetry peak in transit
microlensing events and (ii) the time position of the polarimetry
peak with respect to the photometry peak position in bypass cases.
For simulated microlensing events, we calculate these time shifts
i.e. $\mathrm{\delta_{t}}$. The ratio of this time shift to the mean
value of the time interval between two consecutive polarimetric data
points gives the possible number of polarimetric data points that
can be taken during this time shift. The histogram of this quantity
in the logarithmic scale,
$\mathrm{\log_{10}[\delta_{t}/\bar{T}_{obs}]}$, is plotted in the
right panel of Figure (\ref{fig8}). The number of events which have
$\mathrm{\delta_{t}/\bar{T}_{obs}>3}$ is $\mathrm{221}$, i.e. about
$\mathrm{28.6\%}$ of the total number of simulated events. Detecting
this time shift needs enough polarimetric data points at the
polarimetry peak position(s). We assume that real positions of
photometric peaks can be inferred by fitting process. The number of
simulated events for which FORS2 has observed the polarimetry
peak(s) is $\mathrm{188}$, about $\mathrm{24.3\%}$ of events. On the
whole, $\mathrm{4.7\%}$ of the simulated events have
$\mathrm{\delta_{t}/\bar{\tau}_{p}>3}$ as well as their polarimetry
peaks are covered by synthetic data points, which means that during
the time shift $\mathrm{\delta_{t}}$ more than three polarimetry
data points are taken by FORS2. However, for discerning these
perturbations, the polarimetry residual during this time shift with
respect to the not rotating (spherical) star model should be more
than at least $\mathrm{2\sigma_{p}}$, which is probable according to
fast fall off polarimetry microlensing curves around their peaks.

\begin{figure}
\begin{center}
\includegraphics[angle=0,width=8.cm,clip=]{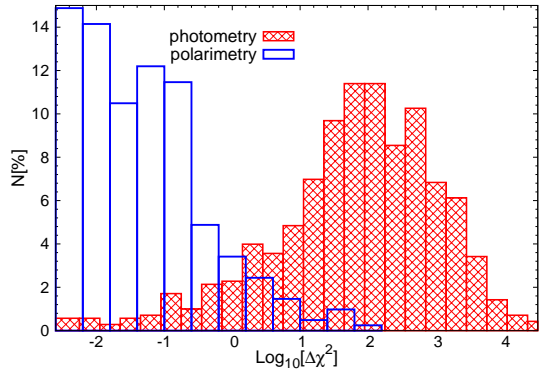}
\caption{The distributions of
$\mathrm{\Delta\chi^{2}=\chi^{2}_{e}-\chi^{2}_{s}}$ in the
logarithmic scale, where $\mathrm{\chi^{2}}$s with the indies
$\mathrm{e}$ and $\mathrm{s}$ are resulted from fitting the real
models of elliptical source stars and the simple models of spherical
source stars respectively, to photometry (red dashed histogram) and
polarimetry data points (blue histogram). For each simulated event,
the parameters used to generate the simple model are similar to the
parameters used to make the real model without considering the
stellar rotation.}\label{fig9}
\end{center}
\end{figure}

(c) Finally, we can discern stellar rotation effects on polarimetry
and photometry microlensing curves by detecting asymmetric
perturbations which are mostly due to gravity-darkening. In that
case, for each simulated event we calculate the
$\mathrm{\Delta\chi^{2}=\chi^{2}_{e}-\chi^{2}_{s}}$ in which
$\mathrm{\chi^{2}_{e}}$ and $\mathrm{\chi^{2}_{s}}$ are resulted
from fitting the real model with the elliptical source star and the
simple model with the not rotating (spherical) source star to the
simulated data points respectively. The parameters of the simple
model are similar to the parameters used to make the real model,
except the source star does not rotate. Also, the radius of the
spherical source star in simple models equals to the radius of the
source star where the lens is entering the source surface, i.e.
$\mathrm{\rho_{\star,s}=\sqrt{u_{0}^{2}+t_{1}^{2}}}$ in transit
microlensing events and $\mathrm{\rho_{\star,s}=\sqrt{a
b}x_{rel}/R_{E}}$ in bypass events. The distributions of
$\mathrm{\Delta\chi^{2}}$ in the logarithmic scale from fitting to
photometry data points (red dashed histogram) and to polarimetry
data points (blue histogram) are plotted in Figure (\ref{fig9}). In
about $\mathrm{83.1}$ and $\mathrm{0.1}$ per cent of simulated light
and polarimetry microlensing curves (respectively), the values of
$\mathrm{\Delta\chi^{2}}$ are higher than $\mathrm{150}$. The small
number of events for which $\mathrm{\Delta\chi_{p}^{2}>150}$ means
that (i) FORS2 by itself can not more likely cover polarimetry
curves of high-magnification microlensing events as well as (ii) its
polarimetry precision is too low to distinguish rotation-induced
perturbations.

Considering all of the mentioned tests for discerning
rotation-induced perturbations, we conclude that in $\mathrm{37}$
and $\mathrm{642}$ simulated events (which contribute $\mathrm{4.8}$
and $\mathrm{83.1}$ per cent of all simulated events) the
polarimetry and photometry perturbations induced by stellar
rotations are distinguishable respectively. Although, the photometry
observation is more efficient than the polarimetry one in detecting
the stellar rotation effects, but by doing only photometry
observation we can not realize the time shift in the photometry peak
position, whereas this time shift is so helpful in discerning the
anomaly kind. However, the small polarimetry efficiency for
detecting stellar rotation effects is rather owing to the lack of
enough number of polarimetry data points to cover the polarimetry
peak(s), if we assume that these observations are done by FORS2 by
itself.

\section{Summary and Conclusions}\label{six}
The stellar rotation causes the ellipticity and gravity-darkening
effects which break the spherical symmetry of the source surface and
the circular symmetry of its surface brightness respectively.
Accordingly, a rotating star has a net polarization signal whose
intensity depends on the inclination angle of the rotational axis as
well as the stellar angular velocity. For fast rotating stars
($\mathrm{\omega>0.6}$) the magnitude of this polarization signal is
comparable with the polarization signal in the transit microlensing
events. This anomaly should be considered in the polarimetry
microlensing calculations of these stars allowing to characterize
the elliptical properties of the source stars. The intrinsic
polarization signals for slow or moderate rotating stars with
$\mathrm{\omega<0.6}$ that are too small to be measured by
themselves, may be highlighted during a high-magnification
microlensing event.

Polarimetry and photometry curves in single-lens microlensing events
of spherical (non rotating) source stars have symmetric shapes with
respect to the time $\mathrm{t_{0}}$ of the closest approach . The
stellar ellipticity breaks this symmetry and causes time shifts
$\mathrm{\delta_{t}}$ in (i) the position of the second polarimetry
peak with respect to the symmetric (with respect $\mathrm{t_{0}}$)
position of the first peak in transit microlensing events and (ii)
the position of the polarimetry peak with respect to the photometry
peak position in bypass cases. However, when
$\mathrm{\xi=0^{\circ}}$, $\mathrm{\xi=90^{\circ}}$ or
$\mathrm{u_{0}=0}$ there is no time shift, because of the symmetric
shape of elliptical source surface projected on the sky plane with
respect to its semimajor and minor axes.

Gravity-darkening produces asymmetric perturbations in polarimetry
and photometry curves of microlensing events. These perturbations
maximize whenever the lens trajectory crosses the projected position
of the stellar rotation pole on the sky plane. In that case, if its
angle with respect to the source semimajor axis equals to
$\mathrm{\xi=90^{\circ}}$ the photometric perturbation maximizes and
if $\mathrm{\xi=\arctan(R_{p}\cos(i)/R_{eq})}$ the polarimetric
perturbation becomes maximum.

The intrinsic polarization signal of an elliptical source is a
degenerate function of the inclination angle and the stellar angular
velocity (see Figure \ref{fig1}). Hence, different elliptical source
stars with the same intrinsic polarization signals can have the same
polarimetry and photometry microlensing curves.

In order to study and compare the photometry and polarimetry
efficiencies for detecting the stellar rotation effects in
high-magnification microlensing events toward the Galactic spiral
arms, we simulated them and considered the fast rotating stars
observed by the Kepler satellite ($\mathrm{\omega>0.1}$) as the
source stars. We generated syntectic (polarimetry and photometry)
data points hypothetically taken by FORS2 and survey and follow-up
telescopes (respectively) for each microlensing event. About
$\mathrm{0.5\%}$ of the source stars had the intrinsic polarization
signals greater than $\mathrm{0.2\%}$ which can potentially be
measured by FORS2 directly. In $\mathrm{4.7\%}$ of total simulated
events more than three polarimetry data points were taken during the
time shift $\mathrm{\delta_{t}}$ by FORS2. In these events, the
rotation-induced time shift and as a result the ellipticity of the
source surface projected on the sky plane can likely be measured. We
also investigated whether asymmetric perturbations in simulated
polarimetry and photometry curves due to the gravity-darkening
effect can be inferred, by calculating the difference between the
values of $\mathrm{\chi^{2}_{e,s}}$ from fitting (a) the real
microlensing models of elliptical source stars and (b) simple
microlensing models of not rotating (spherical) source stars to
synthetic data points. Almost $\mathrm{83.1}$ and $\mathrm{0.1}$ per
cent of photometry and polarimetry curves had
$\mathrm{\Delta\chi^{2}}$ greater than $\mathrm{150}$ respectively.

On the whole, polarimetry and photometry perturbations due to the
stellar rotation were detectable in $\mathrm{4.8}$ and
$\mathrm{83.1}$ per cent of all simulated events respectively.
Therefore, the stellar rotation signatures in high-magnification
microlensing events of early-type stars are mostly realizable
through photometry observations with the technology presently
available. Although, the photometry observation is more efficient
than the polarimetry one for detecting stellar-induced anomalies,
but by only doing photometry observations we can not discern the
time shift in photometry peak positions resulted from the stellar
ellipticity. This time shift is so helpful to realize the nature of
the anomaly and obtain some information about the ellipticity of the
projected source surface. However, the small polarimetry efficiency
for detecting stellar rotation effects is owing to: (i) the low
polarimetry precision of FORS2 and (ii) the long necessary
observational time for taking one polarimetry data point by FORS2
(in comparison with the polarimetry time scales of microlensing
events) which results insufficient number of data points to cover
the polarimetry peak(s). This time shift can likely be distinguished
by high-quality polarimeters of the next generation.

\begin{acknowledgments}
I acknowledge T. Reinhold and S. Bognulo for useful discussions and
comments and also J.-Y., Choi for providing the data of some
high-magnification microlensing events. I thank the referee for
helpful comments and suggestions which significantly improved the
manuscript. Finally, I am grateful to Sh. Baghram and H.Ghodsi for
good comments to improve text of paper. This work was supported by a
grant (94017434) from the Iran National Science Foundation (INSF).
\end{acknowledgments}



\begin{thebibliography}{}
\bibitem[Abney 1877]{Abney1877}
Abney W. D., \ 1877, MNRAS, 37, L278.

\bibitem[Affer et al. 2012]{Affer2012}
Affer L., Micela G., Favata F. \& Flaccomio E., \ 2012, MNRAS, 424,
L11.

\bibitem[Al-Malki et al. 1999]{Almalki1999}
Al-Malki M. B., Simmons J. F. L., Ignace R., Brown J. C., \& Clarke
D., \ 1999, A \& A, 347, L919.

\bibitem[Barnes 2007]{Barnes2007}
Barnes S. A., \ 2007, ApJ, 669, L1167.

\bibitem[Bjorkman \& Bjorkman 1994]{Bjorkman94}
Bjorkman J. E. \& Bjorkman K. S., \ 1994, ApJ, 436, L818.


\bibitem[Bouvier 2013]{Bouvier2013}
Bouvier J., \ 2013, EAS Publications Series, 62, L143.


\bibitem[Borucki et al. 2010]{borucki10}
Borucki, W. J., Koch, D., Basri, G., et al. \ 2010, Sci, 327, L977.

\bibitem[Carciofi \& Bjorkman 2006]{CarciofiB2006}
Carciofi A. C., \& Bjorkman J. E., \ 2006, ApJ, 639, L1081.

\bibitem[Carciofi et al. 2006]{Carciofi2006}
Carciofi A. C., et al., \ 2006, ApJ, 652, L1617.

\bibitem[Casas et al. 2006]{Casas2006}
Casas R., Vaquero J. M., \& Vazquez M., \ 2006, Sol. Phys., 234,
L379.

\bibitem[Chandrasekhar 1960]{Chandrasekhar60}
Chandrasekhar S., \ 1960, Radiative Transfer. Dover Publications,
New York.

\bibitem[Choi et al. 2012]{chio2012}
Choi  J.-Y., Shin I.-G.,  Park S.-Y., et al. \ 2012, ApJ, 751, 41.


\bibitem[Churchwell et al. 2009]{Churchwell2009}
Churchwell E.D., Babler B. L., Meade M. R. et al., \ 2009, PASP,
121, 213.

\bibitem[Collins \& Harrington 1966]{Collins66}
Collins G.W. II, Harrington J. P.  \ 1966, ApJ, 146, L152.

\bibitem[Durney \& Latour 1978]{Durney1978}
Durney B. R., \& Latour J., \ 1978, Geophysical and Astrophysical
Fluid Dynamics, 9, L241.

\bibitem[Ejeta et al. 2012]{Ejeta2012}
Ejeta C., Boehnhardt H., BAgnulo S. \& Tozzi G. P. \ 2012, A \& A,
537, A23.

\bibitem[Fluri \& Stenflo 1999]{Fluri1999}
Fluri D.M., Stenflo J. O., \ 1999, A \& A, 341, L902.

\bibitem[Gallet \& Bouvier 2013]{Gallet2013}
Gallet F. \& Bouvier J., \ 2013, A \& A, 556, L36.

\bibitem[Gaudi \& Haiman 2004]{Gaudi2004}
Gaudi B. S. \& Haiman Z., \ 2004, (arXiv:astro-ph/0401035v1).

\bibitem[Gould 1997]{Gould1997}
Gould A., \ 1997, ApJ, 483, L98.

\bibitem[Harrington \& Collins 1968]{Harrington1968}
Harrington J.P. \& Collins G. W., \ 1968, ApJ, 151, L1051.

\bibitem[Heyrovsk\'y \& Loeb 1997]{Heyrovsky1997}
Heyrovsk\'y D. \& Loeb A., \ 1997, ApJ, 490, L38.

\bibitem[Huber et al. 2014]{Huber2014}
Huber, D., Silva Aguirre, V., Matthews, J. M., et al., \ 2014, ApJS,
211, L2.

\bibitem[Ignace et al. 2009]{Ignace2009}
Ignace R., Al-Malki M.B., Simmons J. F.L., Brown  J. C., Clarke D.
\& Carson J. C., \ 2009, A \& A, 496, L503.


\bibitem[Ingrosso et al. 2012]{Ingrosso12}%
Ingrosso G., Calchi Novati S., De Paolis F., et al. \ 2012, MNRAS,
426, L1496.

\bibitem[Ingrosso et al. 2015]{Ingrosso15}
Ingrosso G., Calchi Novati S., De Paolis F., et al. \ 2015, MNRAS,
446, L1090.

\bibitem[Irwin et al. 2011]{Irwin2011}
Irwin J., Berta Z.K., Burke C.J., et al., \ 2011, ApJ, 727, 56.

\bibitem[Kervella et al. 2004]{Kervella2004}
Kervella P., Th\'evenin F., Di Folco E. \& S\'egransan D., \ 2004, a
\& A, 426, L297.


\bibitem[Kitchatinov 2005]{Kitchatinov2005}
Kitchatinov L. L., \ 2005, Actra Astron, 57, 149.

\bibitem[Koch et al. 2010]{koch10}
Koch, D. G., Borucki, W. J., Basri, G., et al., \ 2010, ApJL, 713,
L79.

\bibitem[Kraft 1970]{Kraft1970}
Kraft, R. P., \ 1970, Spectroscopic Astrophysics. An Assessment of
the Contributions of Otto Struve, L385.

\bibitem[Kroupa et al. 1993]{kroupa93}
Kroupa, P., Tout, C. A., \& Gilmore, G.\ 1993, MNRAS, 262, L545.

\bibitem[Kroupa 2001]{kroupa01}
Kroupa, P.\ 2001, MNRAS, 322, L231.


\bibitem[Le Bouquin et al. 2009]{Bouquin2009}
Le Bouquin J.-B., Absil O., Benisty M., et al., \ 2009, A \& A, 498,
L41.

\bibitem[Maeder \& Meynet 2011]{Maeder2011}
Maeder A. \& Meynet G., \ 2011, arXiv:1109.6171

\bibitem[Maoz \& Gould 1994]{Maoz1994}
Maoz D. \& Gould A., \ 1994, ApJ, 425, L67.


\bibitem[Moniez et al. 2016]{Marc2016}
Moniez M., Rahvar S., Sajadian S., Karami M. \& Ansari R. \ 2016, in
preperation.

\bibitem[Marigo et al. 2008]{Marigo2008}
Marigo P., Girardi L., Bressan A., Groenewegen M. A. T., Silva L.,
Granato G. L., \ 2008, A \& A, 482, L883.


\bibitem[McAlister et al. 2005]{McAlister2005}
McAlister H. A., ten Brummelaar T. A., et al., \ 2005, ApJ, 628,
L439.

\bibitem[McQuillan et al. 2013]{McQuillan2013}
McQuillan A., Aigrain S., \& Mazeh T., \ 2013, MNRAS, 432, L1203.


\bibitem[Lebovitz 1967]{Lebovitz1967}
Lebovitz N.R., \ 1967, ARA \& A, 5, L465.

\bibitem[Ortolani et al. 1995]{Ortolani95}
Ortolani S., Renzini A., Gilmozzi R. et al. \ 1995, Nature, 377,
L701.

\bibitem[Peterson et al. 2004]{Peterson2004}
Peterson D. M., et al., \ 2004, SPIE, 5491, L65.

\bibitem[Rahal et al. 2009]{Rahal2009}
Rahal, Y. R., Afonso C., Albert J.-N., et al. \ 2009, A \& A, 500,
L1027.

\bibitem[Reinhold et al. 2013]{Reinhold2013}
Reinhold T., Reiners A. \& Basri G., \ 2013, A \& A, 560, L4.

\bibitem[Reinhold \& Gizon 2015]{Reinhold2015}
Reinhold T. \& Gizon L., \ 2015, A \& A, .....

\bibitem[Robin et al. 2003]{besancon}
Robin, A. C., Reyl\'{e}, C., Derri\`{e}re, S., Picaud, S., \ 2003, A
\& A, 409, L523.


\bibitem[Russeil 2003]{Russeil2003}
Russeil D., \ 2003, A \& A, 397, 133.

\bibitem[Sajadian 2014]{sajadian12}
Sajadian S., \ 2014, MNRAS, 439, L3007.

\bibitem[Sajadian 2015a]{sajadian15}
Sajadian S., \ 2015a, AJ, 149, L147.

\bibitem[Sajadian 2015b]{sajadian15}
Sajadian S., \ 2015b, MNRAS, 452, L2587.

\bibitem[Schatzman 1962]{Schatzman62}
Schatzman, E., \ 1962, Annales d'Astrophysique, 25, L18.

\bibitem[Scholz \&  Eisl\"{o}ffel 2004, 2005]{Scholz2004}
Scholz A. \& Eisl\"{o}ffel J., \ 2004, A \& A, 419, L249.

\bibitem[Scholz \&  Eisl\"{o}ffel 2005]{Scholz2005}
Scholz A. \& Eisl\"{o}ffel J., \ 2005, A \& A, 429, L1007.

\bibitem[Schneider \& Wagoner 1987]{schneider87}
Schneider P., Wagoner R. V., \ 1987, ApJ, 314, L154.

\bibitem[Simmons et al. 2002]{simmons2002}
Simmons J.F.L., Bjorkman J.E., Ignace R., Coleman I.J., \ 2002,
MNRAS, 336, 501.

\bibitem[Skumanich 1972]{skumanich1972}
Skumanich, A., \ 1972, ApJ, 171, L565.

\bibitem[Stamford \& Watson 1980]{Stamform1980}
Stamford P. A. \& Watson R. D., \ 1980, Acta Astron, 30, L193.


\bibitem[Tinbergen 1996]{Tinbergen96}
Tinbergen J., \ 1996, Astronomical Polarimetry. Cambridge Univ.
Press, New York.

\bibitem[von Zeipel 1924]{vonzeipel}
von Zeipel, H., \ 1924, MNRAS, 84, L665.


\bibitem[Urquhart et al. 2014]{urquhart2014}
Urquhart, J. S., Figura, C. C., Moore, T. J. T., Hoare, M. G., et
al. \ 2014, MNRAS, 437, 1791.


\bibitem[Whitney et al. 2013]{Whitnet2013}
Whitney B.A., et al., \ 2013, ApJS, 207, L30.

\bibitem[Wyrzykowski et al. 2014]{ogle3}
Wyrzykowski, {\L}., et al., 2014, arXiv:1405.3134.

\end{thebibliography}
\end{document}